\documentclass[lettersize,journal, 10pt]{IEEEtran}
\usepackage{xcolor}%for coloring text
\usepackage{graphicx} %for inserting image
\usepackage{multirow}
\usepackage{xcolor,colortbl}
\usepackage{amsmath, amssymb}
\usepackage{hyperref}
\usepackage{caption}
\usepackage{makecell} % For multiline cells
\usepackage{wrapfig}  % For wrapfigure environment

\AtBeginDocument{%
  }
\title{Data-Driven Falsification of Cyber-Physical Systems}

\date{May 2022}

\def \D {\mathcal{D}}
\def \H {\mathcal{H}}
\def \M {\mathcal{M}}

\def \X {\mathcal{X}}

\def \Loc {\mathit{Loc}}
\def \C {\mathcal{C}}
\def \P {\mathcal{P}}

\newtheorem{definition}{Definition}
\newtheorem{problem}{Problem Statement}

\usepackage{algorithmicx}
\usepackage{algorithm}
\usepackage{algpseudocode}
\usepackage{amsmath}
\algdef{SE}[DOWHILE]{Do}{doWhile}{\algorithmicdo}[1]{\algorithmicwhile\ #1}%
\algrenewcommand\algorithmicrequire{\textbf{Input:}}
\algrenewcommand\algorithmicensure{\textbf{Output:}}
\algnewcommand\algorithmicinitialization{\textbf{Initialization:}}
\algnewcommand\Initialization{\item[\algorithmicinitialization]}%

\usepackage{caption}
\usepackage{caption}
\usepackage{subcaption}
\usepackage{adjustbox}
\usepackage{graphics}

\begin{document}

\author{Atanu Kundu, Sauvik Gon, Rajarshi Ray
 \thanks{A. Kundu and S. Gon are students of the Indian Association for the Cultivation of Science (IACS), India. Email: \{ mcsak2346, ugsg2584\}@iacs.res.in. Dr. R. Ray is an Associate Professor at IACS, India. Email: rajarshi.ray@iacs.res.in }}

\maketitle

\begin{abstract}
    Cyber-Physical Systems (CPS) are abundant in safety-critical domains such as healthcare, avionics, and autonomous vehicles. Formal verification of their operational safety is, therefore, of utmost importance. In this paper, we address the falsification problem, where the focus is on searching for an unsafe execution in the system instead of proving their absence. The contribution of this paper is a framework that (a) connects the falsification of CPS with the falsification of deep neural networks (DNNs) and (b) leverages the inherent interpretability of Decision Trees for faster falsification of CPS. This is achieved by: (1) building a surrogate model of the CPS under test, either as a DNN model or a Decision Tree, (2) application of various DNN falsification tools to falsify CPS, and (3) a novel falsification algorithm guided by the explanations of safety violations of the CPS model extracted from its Decision Tree surrogate. The proposed framework has the potential to exploit a repertoire of \emph{adversarial attack} algorithms designed to falsify robustness properties of DNNs, as well as state-of-the-art falsification algorithms for DNNs. Although the presented methodology is applicable to systems that can be executed/simulated in general, we demonstrate its effectiveness, particularly in CPS.  We show that our framework implemented as a tool \textsc{FlexiFal} can detect hard-to-find counterexamples in CPS that have linear and non-linear dynamics. Decision tree-guided falsification shows promising results in efficiently finding multiple counterexamples in the ARCH-COMP 2024 falsification benchmarks~\cite{khandait2024arch}.
\end{abstract}

\begin{IEEEkeywords}
Falsification, CPS, feedforward neural network, Decision Tree, Signal Temporal Logic.
\end{IEEEkeywords}

\section{Introduction}
The traditional simulation and testing techniques can be effective for debugging the early stages of Cyber-Physical-Systems (CPS) design. However, as the design becomes pristine by passing through multiple phases of testing, finding the lurking bugs becomes computationally expensive and challenging by means of simulation and testing alone. Formal verification techniques such as model-checking come in handy here by either proving the absence of bugs in such designs or by providing a \emph{counterexample} behavior that violates the specification. A complementary approach is \emph{falsification}, where the focus is solely on discovering a system behavior that is a counterexample to a given specification. In this work, we address the falsification of safety specifications expressed in signal temporal logic \cite{maler2004monitoring} for CPS given as an executable.

\textbf{Our Contribution} The contribution of this paper is a falsification framework that employs two strategies. First, it connects the falsification of reachability specifications of CPS with the falsification of reachability specifications of deep neural networks (DNNs). Second, it leverages on the inherent interpretability of Decision Trees for faster fastification of signal temporal logic (STL) specifications of CPS. This is achieved by: (1) building a surrogate model of the CPS under test, either as a DNN model or a Decision Tree, (2) application of various DNN falsification tools to falsify CPS, and (3) a falsification algorithm guided by the interpretations of safety violation of a CPS model extracted from its Decision Tree surrogate.  The proposed framework has the potential to exploit a repertoire of \emph{adversarial attack} algorithms designed to falsify the robustness properties of DNNs as well as state-of-the-art falsification algorithms designed for DNNs. A reachability specification of a CPS directly maps to a reachability specification of a DNN when the latter approximates the former. Furthermore, a recent work has shown that a reachability specification of a DNN can be reduced to an equivalid set of robustness specifications \cite{shriver2021reducing}. A robustness specification conveys that the network's output on an $\epsilon$-perturbed input matches with its output on the unperturbed version \cite{DBLP:conf/sp/Carlini017, DBLP:journals/corr/SzegedyZSBEGF13, yuan2019adversarial}. This reduction allows the use of adversarial attack algorithms in the literature of DNN \cite{goodfellow2014explaining, DBLP:conf/iclr/KurakinGB17a, DBLP:conf/cvpr/Moosavi-Dezfooli16, madry2017towards}, which are targeted to falsify robustness specification, for falsification of CPS. Our framework provides an interface to several tools that can falsify the safety properties of DNNs as well as an interface to a variety of \emph{adversarial attack} algorithms that are aimed towards the falsification of robustness specification of DNNs. These tools and algorithms have complementary strengths. As a result, there is a high likelihood of success in a falsification task when trying the various falsification tools and algorithms through the interface. A limitation of DNN-based falsification is the considerable computational resources it consumes to build a surrogate model from data and consequently to find a counterexample. To address this limitation, our framework integrates a falsification strategy based on building a decision tree as a surrogate model. A decision tree, in contrast to a DNN, is comparatively cheaper to build and its inherent explainability provides a mechanism for efficiently searching multiple counterexamples which we shall discuss in detail later in the paper. The primary challenge of this framework is to construct a faithful surrogate model of the CPS, which we carry out by generating traces from the CPS executable and learning (supervised) from the traces. For complex CPS such as highly autonomous vehicles, designing a faithful model such as a hybrid automaton \cite{Alur92hybridautomata} may not be practical due to complex dynamics. Furthermore, models combining explicit hybrid automaton and \emph{black-box} executable may be relevant and practical for real-world CPS \cite{FanQM017}. The proposed framework can  be beneficial for falsifying these types of system, since it relies only on CPS trajectory data. Although the proposed falsification framework applies to executable systems in general, this work is focused on exploring its effectiveness in CPS in particular, and we report our findings.
In summary, the main contributions of the paper are as follows.
\begin{itemize}
    \item A data-driven framework for falsification of CPS safety specifications given as an executable (black-box). A tool \textsc{FlexiFal} implementing  the framework.
    \item A labeled dataset representing a number of CPS obtained from their executed/simulated trajectories. The dataset can be beneficial for building machine learning models for CPS analysis and is publicly available. \footnote{\href{https://drive.google.com/drive/u/3/folders/12PzBA6dGymynYH4_Ycn_91vIqpQrogsc}{Data sets of the CPS under test}.}
    \item Neural network surrogates of representative hybrid automata and Simulink models of CPS are made publicly available in onnx format. \footnote{\href{https://gitlab.com/Atanukundu/NNFal/-/tree/main/network}{Neural-network models of the corresponding CPS.}}
\end{itemize}

\section{Preliminaries}

\noindent In this work, we use a fully connected feed-forward neural network with the rectified linear unit (\emph{ReLU}) activation function, and a decision tree as the machine learning models to approximate a CPS. % 
A neural network can be seen as computing a function $\textit{N}: \mathbb{R}^{n_1} \rightarrow \mathbb{R}^{n_2}$, where $n_1$ and $n_2$ denotes the number of inputs and outputs of the network respectively. A decision tree is a supervised machine learning model for classification and regression tasks. 
In this work, we employ CART (Classification and Regression Tasks) among the several decision tree construction algorithms~\cite{breiman2017classification}. Safety requirements are often represented using \emph{reachability specification}. A reachability specification comprises a set of initial configurations ($\mathcal{I}$) and a set of unsafe configurations ($\mathcal{U}$) of the system and specifies that for all inputs $x \in \mathcal{I}$, the output of the system does not belong to $\mathcal{U}$. The falsification problem of a reachability specification of a network is given as:
 \begin{definition}[Falsification of Reachability Specification]
     Given a deep neural network that computes the function $\textit{N}: \mathbb{R}^{n_1} \rightarrow \mathbb{R}^{n_2}$ and a reachability specification ($\mathcal{I}$, $\mathcal{U}$) such that $\mathcal{I} \subseteq \mathbb{R}^{n_1}$ and $\mathcal{U} \subseteq \mathbb{R}^{n_2}$, the falsification problem is to find $x \in \mathcal{I}$ such that $\textit{N}(x) \in \mathcal{U}$. 
 \end{definition}
  
\noindent The robustness specification when applied to classification models refers to the network's ability to keep its classification unchanged when its input is perturbed within an $\epsilon$ neighborhood. 
We formally define a local robustness specification of DNN as follows:
 \begin{definition}[Local Robustness Specification]
     Given a deep neural network that computes the function $\textit{N}: \mathbb{R}^{n_1} \rightarrow \mathbb{R}^{n_2}$ and an input $x \in \mathbb{R}^{n_1}$, the network is said to be locally robust with respect to the input $x$ if $\textit{N}(x) = \textit{N}(x')$ for all $x'$ in the $\epsilon$ neighbourhood of input $x$.
 \end{definition}
\noindent The robustness of a network can be refuted by showing the presence of an $\epsilon$-perturbed input for the given input $x$ such that the network's outcome on $x$ and the perturbed version do not match. 

\noindent In the text, we depict an $n$-dimensional deterministic CPS by a function $\M:\P \times \mathbb{C}(\mathbb{R}) \times \mathbb{R} \to \mathbb{R}^n$, where $\P \subseteq \mathbb{R}^n$ is the set of initial system configurations, $\mathbb{C}(\mathbb{R})$ is the space of continuous functions $u: \mathbb{R} \to \mathbb{R}^m$ representing a time-varying $m$-dimensional input signal to the CPS. An input signal is a function that maps a time instant $t$, to an input $u(t) \in \mathbb{R}^m$ to the CPS. The function $\M(x_0, u, t)$ represents the state that the CPS reaches, starting from the initial system configuration $x_0$, under the influence of the input signal $u$ and running for $t$ time units. In this work, we restrict ourselves to deterministic CPS. A CPS trajectory depicts how the variables in the system change over time. It can be formally defined as follows.

\begin{definition}
A trajectory of a CPS $\M:\P \times \mathbb{C}(\mathbb{R}) \times \mathbb{R} \to \mathbb{R}^n$, starting from an initial system configuration $x_0 \in \P$, with an input signal $u \in \mathbb{C}(\mathbb{R})$ over a time horizon $T$ is given by a function $\Gamma: [0, T] \to \mathbb{R}^n$ such that for any $t_1$, $t_2$ $\in [0, T]$ where $t_1 < t_2$, $\Gamma(t_2) = \M(\Gamma(t_1),u,t_2 - t_1)$. 
\end{definition}

\noindent We now state a CPS's falsification problem, which we address in Section \ref{nnfal} of the paper. A general falsification problem of STL specification is defined and addressed in Section \ref{dtfal}.
\begin{problem}
Given a CPS $\M:\P \times \mathbb{C}(\mathbb{R}) \times \mathbb{R} \to \mathbb{R}^n$ and a reachability specification ($\mathcal{I}$, $\mathcal{U}$) where $\mathcal{I} \subseteq \P \times \mathbb{C} (\mathbb{R}) \times \mathbb{R}$ and $\mathcal{U} \subseteq \mathbb{R}^n$, find a tuple $\C = (x_c, u_c, t_c)$, an element of $\mathcal{I}$ with an initial system configuration $x_c$ , an input signal $u_c$ and a time $t_c$ such that $\M(x_0,u_c,t_c) \cap \mathcal{U} \ne \emptyset$. 
\end{problem}

\noindent Such a tuple is called a counterexample. In this work, we are limited to finding a counterexample having a piecewise-constant input signal, with equally spaced control points over time. Bounded ranges for each dimension of the input are assumed to be known.

\section{Framework for CPS Falsification}\label{nnfal}
\begin{figure}
    \centering
    \includegraphics[scale=0.32]{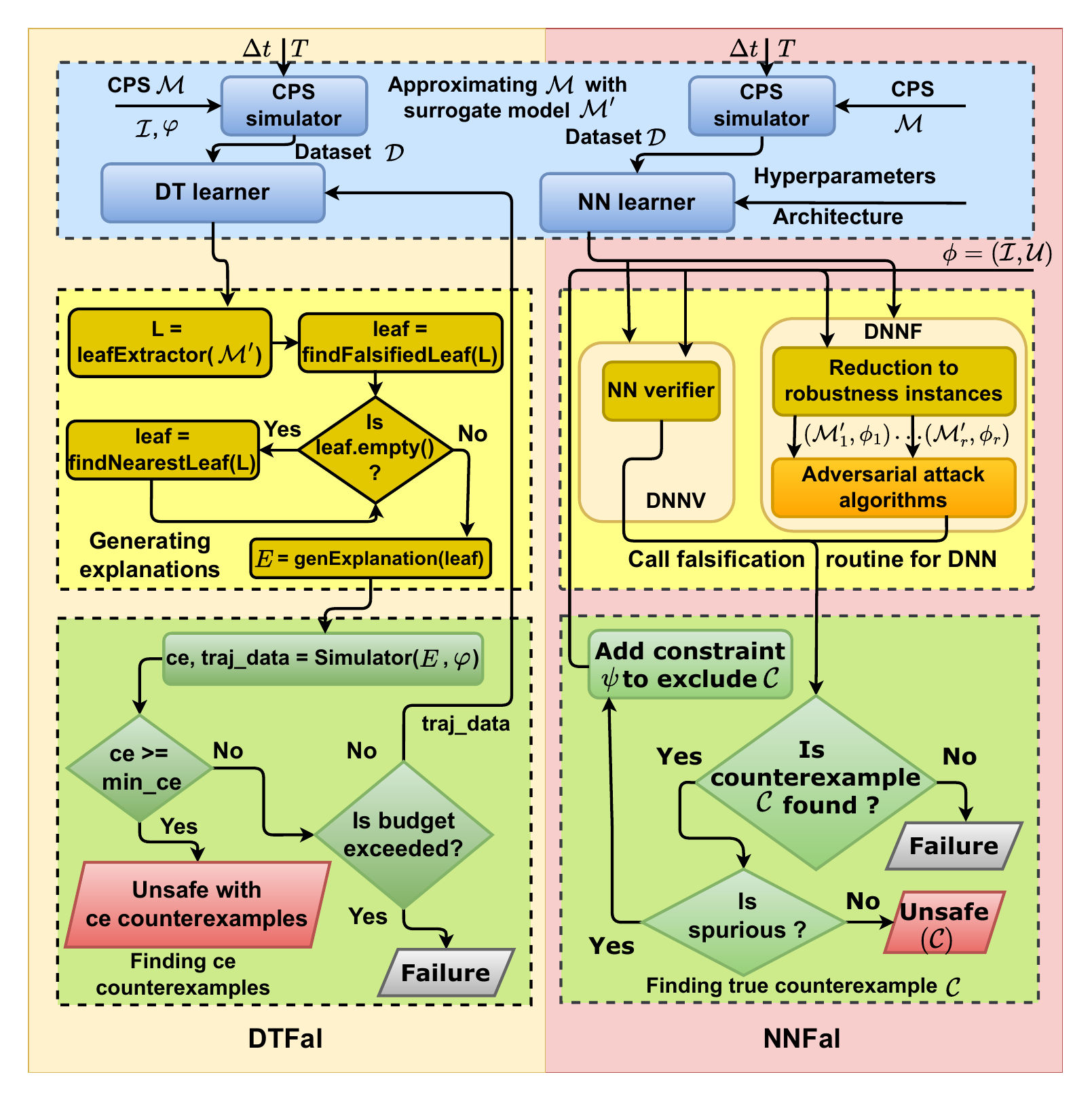}
    \caption{The flowchart of the falsification framework \textsc{FlexiFal}. The framework incorporates two distinct falsification strategies, \textsc{DTFal} (left) and \textsc{NNFal} (right).}
    \label{fig:proposed_arch}
\end{figure}

 \noindent We now present our falsification framework \textsc{FlexiFal}, which consists of mainly two procedures, DNN-based falsification and falsification using a Decision Tree. This section focuses on describing the DNN-based falsification \textsc{NNFal}, while the decision tree-based falsification procedure is described in the subsequent section. The foremost step is building a DNN model from the CPS (given as an executable) that we intend to falsify against a given reachability specification. This construction reduces the falsification problem of the CPS to the falsification of the constructed DNN. The next step is to search for a counterexample of the reachability specification in the constructed DNN either by using one of the DNN falsification tools or by using one of the adversarial attack algorithms to falsify local robustness specification. The relation between the falsification of a reachability specification and the falsification of the local robustness specification of DNN is shown in~\cite{shriver2021reducing}. The last step of our algorithm is to validate whether the generated counterexample is spurious or real by execution on the actual CPS. If the execution lends the counterexample as spurious, we modify the reachability specification by adding necessary constraint(s) in order to exclude the spurious counterexample from the search space. The falsification tool is invoked with the modified specification of the DNN. The modified specification forbids the tool from repeatedly generating spurious counterexamples. The steps of invocation, validation, and specification modification of the falsification tool are repeated. The algorithm terminates on two conditions. The first condition is when a valid counterexample is found, declaring that the DNN, and thus the CPS is falsified. The other termination condition is due to a timeout before finding a valid counterexample, in which case, the algorithm terminates by declaring failure. Figure \ref{fig:proposed_arch} depicts an overview of our proposed framework, and the block diagram of \textsc{NNFal} is shown on the right. We now discuss the details of \textsc{NNFal}.

\subsection{CPS design to Neural Networks}

\subsubsection{Dataset Generation from CPS} \label{cpstodataset}
The primary step in constructing a neural network model from the CPS is building a labeled dataset. Since trajectories of a CPS represent its behavior, a machine learning model ought to be learned from trajectories. In this work, we consider models of CPS (Simulink/Hybrid automata) and then use model simulators (Matlab Simulink model simulator/ XSpeed hybrid automata simulator) to generate a dataset $\D$ from a CPS $\M$. In this work, we use a model solely to generate simulated trajectories. The simulators use numerical ode solvers such as CVODE \cite{serban2005cvodes} to simulate continuous behaviors of the CPS. To formally describe the dataset, we use some notations as follows. We represent a time discretization of a trajectory of the CPS at $\Delta t$ time-steps as $\Gamma (j \Delta t) = \M (x_0, u, j \Delta t)$. To represent the time-varying input signal $u(t)$ in the dataset, we consider $k$ partitions of the time horizon and consider a fixed input in each interval. This scheme discretizes the time-varying input into a piece-wise constant input, which can be represented as an element of $\mathbb{R}^{k \times m}$. The larger the partitions $k$, the closer the data representation becomes to the actual time-varying input. However, increased partitions increase the parameters in the dataset and hence the data size. In this work, we determinize the \emph{may} transition semantics and the mode-switching updates. To determinize the \emph{may} transition semantics, we generate our data with urgent transition semantics in the CPS simulator, which means a transition is taken as soon as its guard is enabled. To address the uncertainty in the state updates, we remove non-deterministic choices in the CPS across all simulations. %Note that data generation with these discussed assumptions limits the set of possible behaviors of the CPS shown by the simulator. This in turn affects the quality of the machine learning model that we learn to approximate the CPS and remains a limitation of this work.
Algorithm \ref{algo:dataset_generations} depicts the generation of the dataset $\D$ for $N$ simulated trajectories. The algorithm accepts as inputs the CPS $\M:\P \times \mathbb{C}(\mathbb{R}) \times \mathbb{R} \to \mathbb{R}^n$, set of initial configuration $ \subseteq \P$, input space given by the lower and upper bound on each input, the input discretization constant $k$, time-horizon $T$, trajectory discretization time-step $\Delta t$, and the number of trajectories $N$ to be generated. Each trajectory is generated for a randomly selected initial configuration $x_0 \in \mathcal{I}$ (Line~\ref{initial}) and a randomly created input signal (lines~\ref{input_start} - \ref{end_loop}). For creating $k$ equally spaced piecewise constant input signal, the loop in line~\ref{input_start} iterates over $k$, and the inner loop (lines~\ref{n_dim} - \ref{end_inner_loop}) generates a random input ($u_j$) using the ranges of each input. Line~\ref{eq_range} depicts the construction of a constant input signal $u_j$ in the time interval [$j.\frac{T}{k}$, $(j+1).\frac{T}{k}$]. The for loop in line~\ref{time_discrete} is iterated over the trajectory discretization time points ($\lceil T/\Delta t\rceil$ many) to compute $\Gamma (j \Delta t)$. Line~\ref{data_formation} makes a datapoint by concatenating $x_0$, $u$, $j \Delta t$, and $\Gamma (j \Delta t)$ and storing in the dataset $\D$.
The input to the CPS, $x_0$, $u$, and $t$ are the input features, and the corresponding output $\Gamma (j \Delta t) = \M (x_0, u, t)$ is stored as labeled output in the dataset $\D$ shown in Table \ref{tab:dataset_structure}. This dataset is used to learn the CPS behavior in a feed-forward deep neural network model that approximates the function $\M (x_0, u, t)$.

\begin{algorithm}[ht]
    \caption{Dataset generation from CPS.}
    \footnotesize
    \label{algo:dataset_generations}
    \begin{algorithmic}[1]
        \Require The CPS $\M:\P \times \mathbb{C}(\mathbb{R}) \times \mathbb{R} \to \mathbb{R}^n$, the set of initial system configurations $\subseteq \P$, the input space given as a hyper-rectangle [$u_1^{lw}$, $u_1^{up}$] $\times$ [$u_2^{lw}$, $u_2^{up}$] $\times \ldots \times$ [$u_p^{lw}$, $u_p^{up}$], the input discretization $k$, the \emph{time-horizon} $T \in \mathbb{R}$, the \emph{time-step} $\Delta t$, and the number of trajectories to be generated $N$.
        \Ensure Dataset $\D$ stores the time discretization of $N$ trajectories.
        \Initialization $\Gamma (k) \gets \emptyset$, $\D \gets \emptyset$.
        \For{$i = 1$ to $N$} \label{algo:n_traj}\Comment{Generate $N$ trajectories.}
            \For{$j=0$ to $k-1$} \label{input_start}\Comment{Generate a piecewise constant input signal with $k$ equally spaced inputs.}
             \For{$r = 1$ to $p$} \label{n_dim}\Comment{Select a random input.}
                \State $u_j[r]$ = random([$u_r^{lw}$, $u_r^{up}$])\label{k_spaced_ip}
             \EndFor \label{end_inner_loop}
             \State $u$([$j.\frac{T}{k}$, $(j+1).\frac{T}{k}$]) = $u_j$ \Comment{Construct constant input signal for the intervals [$0$,$\frac{T}{k}$],\dots,[$(k-1).\frac{T}{k}$, $k.\frac{T}{k}$].}\label{eq_range}
            \EndFor \label{end_loop}
           \State Pick an initial configuration $x_0 \in \mathcal{I}$ \label{initial}
            \For{$j = 0$ to $\lceil T/\Delta t\rceil$}\label{time_discrete}
                \State $\Gamma (j \Delta t) = \M(x_0, u, j \Delta t)$\label{trajctory}
            %\EndFor
             %\For{$k = 0$ to len\big($\Gamma (k)$\big)}
                \State $Datapoint$ =  $\langle x_0, [u_0,u_1,\ldots,u_{k-1}], j \Delta t, \Gamma (j \Delta t)\rangle$\label{data_formation}
                \State $\D.append(Datapoint)$
             \EndFor
        \EndFor
        \State \Return $\D$
    \end{algorithmic}
\end{algorithm}

\begin{table}[h]
    \caption{The dataset $\D$ of N trajectories. Each row is treated as labeled data for DNN construction by supervised learning.% Initial is the initial system configuration.
    }
    \footnotesize
    \label{tab:dataset_structure}
    \centering
    \begin{tabular}{cccc}
       Initial  & Input signal & Time & Output \\
       \hline
        $\langle x_0^1$ $\rangle$ &  $\langle u_0^1,\dots, u_{k-1}^1\rangle$ & $0$ & $\langle \M(x_0^1, u^1, 0)\rangle$ \\
        $\langle x_0^1\rangle$ & $\langle  u_0^1,\dots, u_{k-1}^1\rangle$ & $\Delta t$ & $\langle \M(x_0^1, u^1, \Delta t)\rangle$ \\
        $\dots$ & $\dots$ & $\dots$ \\
        $\langle x_0^1\rangle$ & $\langle  u_0^1,\dots, u_{k-1}^1\rangle$ & $j\Delta t$ & $\langle \M(x_0^1, u^1, j\Delta t)\rangle$ \\
        $\dots$ & $\dots$ & $\dots$ \\
        $\dots$ & $\dots$ & $\dots$ \\
        $\langle x_0^N\rangle$ & $\langle  u_0^N,\dots, u_{k-1}^N\rangle$ & $0$ & $\langle \M(x_0^N, u^N, 0)\rangle$ \\
        $\dots$ & $\dots$ & $\dots$ \\
        $\langle x_0^N\rangle$ & $\langle  u_0^N,\dots, u_{k-1}^N\rangle$ & $j\Delta t$ & $\langle \M(x_0^N, u^N, j\Delta t)\rangle$ \\
    \end{tabular}
\end{table}

\subsubsection{Data Pre-processing}
Feature scaling is a technique used in machine learning to normalize the range of input features in a dataset. It is necessary for the machine learning algorithms to interpret all the features on the same scale. If the input features have largely varying ranges, it can lead to some features dominating the learning process, causing other features to be ignored. We apply MinMax scaling, which scales the input features in a range of 0 to 1 as follows: $\hat{x}_{l,m} = \frac{x_{l,m} - min_m}{max_m - min_m}$,
where $x_{l,m}$ denotes the value of feature $m$ of record $l$ and $\hat{x}_{l,m}$ is the corresponding normalized value. Here, $min_m$ and $max_m$ denote the minimum and maximum values of the $m$th input feature, respectively. 

\subsubsection{Neural Network Training}

\begin{table*}[h]
    \caption{The table shows the details of neural networks learned from the CPS trajectories. \textit{\#Network Layers} is the number of hidden layers in the neural network. \textit{Architecture} shows the number of neurons in each hidden layer. Hyperparameters show- the epochs and the batch size.
    %Early stopping is not used on Nav, Two Tanks, and AT CPS.
    \textit{Dataset Size} shows the number of data entries (in Millions) in the dataset. \textit{Learning Time} shows the time taken to train the network. \textit{Dataset Generation Time} is the time to produce the dataset from the CPS. 
    %This is to evaluate the effect of the network architecture on the performance of falsification.
    }
    \centering
    %\footnotesize
     \begin{adjustbox}{width=1\textwidth}
     %\resizebox{\columnwidth}{!}{
     \begin{tabular}{|c|c|c|c|c|c|c|c|}
        \hline
        \multirow{2}{*}{CPS} & \multirow{2}{*}{\#Network Layers} & \multirow{2}{*}{Architecture} & \multicolumn{2}{c|}{Hyperparameters} & Dataset Size & Learning Time & Dataset Generation  \\
        \cline{4-6}
        & & & Epochs & Batch-Size  & (Millions)  & (Hours) & Time (Secs) \\ 
        \hline
       %Oscillator & Osc\_NN & 12 & $\underbrace{1024 \times \dots \times 1024}_{\text{layer 1} \to \text{layer 5}} \times \underbrace{512 \times \dots \times 512}_{\text{layer 6} \to \text{layer 9}} \times \underbrace{256 \times \dots \times 256}_{\text{layer 10} \to \text{layer 12}}$ & %activation="relu", regularizer=regularizers.L2(0.005), optimizer="adam",loss="MSE", epochs=200, batch\_size=256 \{relu, adam, MSE, 200, 256\} and L2 Regularization\\
       Oscillator & 6 & 512 $\times$ 256 $\times$ 128 $\times$ 64 $\times$ 32 $\times$ 16 & 16 & 256  & 6.1  & 0.3 & 79\\
        \hline
        %Two tanks & Two tanks\_NN-1 & 9 & $\underbrace{512 \times \dots \times 512}_{ \text{layer 1} \to \text{layer 9}}$ &\{relu, adam, MSE, 200, 256\} \\
        Two tanks & 5 & 512 $\times$ 256 $\times$ 128 $\times$ 64 $\times$ 32 & 200 & 256 & 8.2 & 4.3 & 71\\
        \hline
        %Two tanks & Two tanks\_NN-2 & 6 & 512 \times 128 \times 128 \times 64 \times 64 \times 32 & \{relu, adam, MSE, 110, 256\} \\
        %\hline
        %NAV\_30 & NAV\_30\_NN-2 & 4 & 164 $\times$ 128 $\times$ 64 $\times$ 32 & \{relu, adam, MSE, 200, 512\}\\
        Navigation & 6 & 512 $\times$ 256 $\times$ 128 $\times$ 64 $\times$ 32 $\times$ 16 & 200 & 256 & 11.8 & 3.6 & 318\\
        \hline
        Bouncing ball & 7 & 512 $\times$ 256 $\times$ 256 $\times$ 128 $\times$ 128 $\times$ 64 $\times$ 32 & 21 & 256 & 7.2  & 0.7 & 36 \\
        \hline
        ACC & 5 & 512 $\times$ 256 $\times$ 128 $\times$ 64 $\times$ 32 & 19 & 256 & 6.5 & 0.5 & 98 \\
        \hline
        AGCAS & 5 & 512 $\times$ 256 $\times$ 128 $\times$ 64 $\times$ 32 & 54 & 128 & 3.5  & 1.4 & 1622\\
        \hline
        CC & 5 &  512 $\times$ 256 $\times$ 128 $\times$ 64 $\times$ 32 & 7 & 256 & 33.3  & 0.7 & 743\\
        \hline
        AT & 5 & 512 $\times$ 256 $\times$ 128 $\times$ 64 $\times$ 32 & 150 & 256 & 10  & 4 & 1251 \\
        \hline
        %  SC & 5 & 512 $\times$ 256 $\times$ 128 $\times$ 64 $\times$ 32 & 19 & 256 & 17.5 & 0.9 & 343\\
        % \hline
        AFC & 5 & 512 $\times$ 256 $\times$ 128 $\times$ 64 $\times$ 32 & 30 & 256 & 33.6  & 2.3 & 3022\\
        \hline
        SB1 & 5 & 512 $\times$ 256 $\times$ 128 $\times$ 64 $\times$ 32 & 200 & 16 & 0.05 & 0.1 & 151.74\\
        \hline
    \end{tabular}
   % }
    \end{adjustbox}
    \label{nn_training}
\end{table*}

\noindent The design choice of the approximating neural network has been made following the principle of \emph{Ockham's razor}. We learn a fully connected feed-forward neural network. For simplifying the design space exploration, the choices in the hyperparameter settings have been kept fixed, whereas the choices in the network graph structure have been explored on a trial-and-error basis. 

\noindent \textbf{Hyperparameter setting:} The networks have been trained with \emph{ReLU} activation function, using the Adam optimizer with a learning rate of $\lambda = 0.0001$, and on the mean squared error (MSE) loss function.

\noindent \textbf{Network Graph Structure} We start with a simple structure having a few hidden layers and a few nodes per layer. We gradually increase the architecture complexity by trial and error until the obtained network's performance is satisfactory. A network's performance is assessed against a fixed set of falsification instances according to our methodology using \emph{falsification rate} as the metric.

Given a falsification instance, the \emph{falsification rate} of a network is the number of times our framework successfully finds a counterexample out of a given number of runs using the network. A higher \emph{falsification rate} suggests a greater likelihood of finding a counterexample using the network.

\noindent \textbf{Training with Cross-Validation:} We use the early-stopping algorithm that monitors the validation loss in order to stop the training before the model overfits. A \emph{patience} value of 5 has been used. The early-stopping algorithm stops the learning iterations when the validation loss does not improve over the last \emph{patience} many epochs. The network with the lowest validation loss over the epochs is saved. For some of the CPS (Navigation, Two-tanks, AT, and SB1), we did not use early stopping and instead trained for a fixed number of epochs, monitoring the training loss and saving the model with the lowest training loss instead.
The architecture of the chosen neural network for each CPS and the hyperparameter settings used in the training process are detailed in Table \ref{nn_training}.

\subsection{Generating a Counterexample}
\noindent Once a network $\M'$ is built using the dataset $\D$, the next step of our algorithm is to generate a counterexample to the given reachability specification. Our framework provides an interface to many adversarial attack algorithms and DNN verification tools which can be invoked to falsify the given reachability specification on the network $\M'$. In particular, it uses \emph{deep neural network falsifier} (DNNF) \cite{shriver2021reducing}, a falsification framework that reduces a reachability specification to a set of equivalent local robustness specifications. This reduction opens up the applicability of a rich class of adversarial attack algorithms such as the Fast Gradient Sign Method (\emph{fgsm}), Basic Iterative Method (\emph{bim}), DeepFool, and Projected Gradient Descent (\emph{pgd}) for falsification.
In addition, our framework provides an interface to DNN verification tools via \emph{deep neural network verifier} (DNNV)~\cite{DBLP:conf/cav/ShriverED21}. DNNV supports verification tools such as Reluplex, Neurify, and Nnenum and automatically translates the given network and property to the input format of the invoked verifier. 
\subsection{Eliminating Spurious Counterexamples}
\noindent We determine whether the counterexample given by a DNN falsifier is spurious or real by simulating it in the actual CPS model. We use the simulation engine in \textsc{XSpeed} for CPS given as hybrid automata, and we use MATLAB for simulating CPS given as Simulink models. The simulation engine computes a trajectory from the counterexample $\C = (x_c, u_c, t_c)$ and checks whether the trajectory intersects with the set of unsafe configurations $\mathcal{U}$. If the trajectory does not intersect with $\mathcal{U}$, we mark the counterexample as spurious. Each spurious counterexample is stored in a set $\psi \gets \psi \cup \C$. We remove $\psi$ from the subsequent falsification problem by modifying the property as ($\mathcal{I'},\mathcal{U}$), where $\mathcal{I'} = \mathcal{I} - \psi$. The updated property is evaluated on the falsifier again to find a counterexample, and this process is continued until a true counterexample is found within the time and memory limit.

\section{A Falsification Procedure using Decision-Tree}\label{dtfal}
Training networks can be expensive. As an alternative, we propose a falsification procedure guided by a decision tree surrogate model of the CPS. Unlike \textsc{NNFal} which is restricted to falsify reachability specification, this procedure can falsify generic Signal Temporal Logic (STL) specifications. We briefly present STL, a specification language to express the properties of CPS \cite{maler2004monitoring}. 
The syntax of an STL formula \( \varphi \) over a finite set of real valued variables  $V$ is defined by the following grammar:
$$\varphi ::= v \sim d \mid \neg \varphi \mid \varphi_1 \land \varphi_2 \mid \varphi_1 \, U_I \, \varphi_2 $$
where \( v \in V \), \( \sim \in \{<, \leq\} \), \( d \in \mathbb{Q} \) is a Rational, and \( I \subseteq \mathbb{R}^+ \) is an interval. The semantics of STL is given with respect to a \emph{signal}, which in the context of CPS is a trajectory $\Gamma$. $\Gamma_v$ denotes the projection of $\Gamma$ on $v\in V$.

\begin{definition}[STL Semantics~\cite{maler2004monitoring}]
    We denote $(\Gamma, t) \models \varphi$ to mean that a finite trajectory $\Gamma: [0, T] \to \mathbb{R}^n$ satisfies the STL property $\varphi$ at the time point $t \in [0, T]$. The rules of satisfaction are as follows: 

\begin{align}
    (\Gamma, t) &\models v \sim d  
        &&\Leftrightarrow \quad \Gamma_v (t) \sim d \\
    (\Gamma, t) &\models \neg \varphi  
        &&\Leftrightarrow \quad (\Gamma, t) \not\models \varphi \\
    (\Gamma, t) &\models \varphi_1 \land \varphi_2  
        &&\Leftrightarrow \quad (\Gamma, t) \models \varphi_1 \text{ and } (\Gamma, t) \models \varphi_2 \\
    (\Gamma, t) &\models \varphi_1 \, U_I \, \varphi_2  
        &&\Leftrightarrow \quad 
        \begin{aligned}[t]
            &\exists t' \in (t + I) \cap T : (\Gamma, t') \models \varphi_2 \\
            &\text{and } \forall t'' \in (t, t'), (\Gamma, t'') \models \varphi_1
        \end{aligned}
\end{align}

\end{definition}

The temporal operators \emph{always} ($\Box_I$) and \emph{eventually} ($\Diamond_I$) are defined using the \emph{until} operator as:
\[
\Box_I \varphi = \neg(\top \, U_I \, \neg \varphi), \quad
\Diamond_I \varphi = \top \, U_I \, \varphi.
\]

Informally, $\Box_{(a,b)} \varphi$ is true if $\varphi$ holds at all times within the interval $(a,b)$,  whereas $\Diamond_{(a,b)} \varphi$ is true if $\varphi$ holds at some time within $(a,b)$. 

\subsection{Robustness of a Trajectory}

The notion of \emph{robustness}~\cite{Donze2010Robust} of a trajectory to a given STL specification evaluates how "closely" the trajectory satisfies a specification. Mathematically, it is a real-valued function $\rho(\varphi; \Gamma, t) \in \mathbb{R}$, which measures the degree of satisfaction of $\varphi$ by $\Gamma$ at time point $t$. The notion of satisfaction of a specification is lifted from a yes/no answer to a quantitative measure. We utilize the notion of robustness of CPS trajectories to classify them using a decision tree based on their "closeness" to violating a given STL safety specification. A formal description of robustness is given as follows.   

\[
(\Gamma, t) \models \varphi \;\iff\; \rho(\varphi; \Gamma, t) \geq 0,
\]
\[
(\Gamma, t) \not\models \varphi \;\iff\; \rho(\varphi; \Gamma, t) < 0.
\]
For atomic predicates $v \sim d$, the robustness is defined as: 
\[
\rho(v \sim d; \Gamma, t) = d - \Gamma_{v}(t).
\]
The robustness of complex STL formulas is defined as:

\begin{align}
    \rho(\neg \varphi; \Gamma, t) &= -\rho(\varphi; \Gamma, t), \\
    \rho(\varphi_1 \land \varphi_2; \Gamma, t) &= 
    \min\left(\rho(\varphi_1; \Gamma, t), 
              \rho(\varphi_2; \Gamma, t)\right), \\
    \rho(\varphi_1 \, U_I \, \varphi_2; \Gamma, t) &=
    \max_{t' \in (t + I) \cap [0,T]} 
    \big( \min\big( \rho(\varphi_2; \Gamma, t'), \nonumber \\
    &\quad \inf_{t'' \in [t, t')} 
    \rho(\varphi_1; \Gamma, t'') \big) \big).
\end{align}

The robustness of derived operators are defined as follows:
\begin{align*}
\rho(\Diamond_I \varphi; \Gamma, t) &= \max_{t' \in (t + I) \cap [0,T]} \rho(\varphi; \Gamma, t'), \\
\rho(\Box_I \varphi; \Gamma, t) &= \min_{t' \in (t + I) \cap [0,T]} \rho(\varphi; \Gamma, t').
\end{align*}

\noindent We mention $\rho(\varphi; \Gamma)$ to mention the robustness of the trajectory at time $t = 0$, that is $\rho(\varphi; \Gamma, 0)$. We now state the CPS falsification problem addressed in this section. 
\begin{problem}
Given a CPS $\M:\P \times \mathbb{C}(\mathbb{R}) \times \mathbb{R} \to \mathbb{R}^n$, an initial set of configurations $\mathcal{I} \subseteq \P \times \mathbb{C} (\mathbb{R}) \times \mathbb{R}$ and a STL safety specification $\varphi$, find a tuple $\C = (x_c, u_c, t_c)$, an element of $\mathcal{I}$ such that the trajectory $\Gamma: [0, t_c] \to \mathbb{R}^n$ of $\M$ starting from an initial configuration $x_c$, with an input signal $u_c$ over a time horizon $t_c$ has robustness $\rho(\varphi; \Gamma) < 0$.  
%$\M(x_0,u_c,t_c) \cap \mathcal{U} \ne \emptyset$.  %We refer to such a by counterexample by $\C = (x_c, u_c, t_c)$. 
\end{problem}

\noindent The first step is to construct the dataset using a CPS simulator in a similar way we generated the dataset for building a DNN, but now using the robustness of CPS trajectories as the output feature. A decision tree ($\M'$) is learned from the generated data set. The decision tree is then traversed to find the leaf nodes that classify trajectories violating the given safety specification, that is with robustness less than 0. If there is no such leaf node, then our algorithm finds the leaf node that characterizes the trajectories \emph{closest} to violating the safety specification. An explanation for violating the safety specification is then extracted by backtracking the path from the identified \emph{leaf} node to the root and conjoining the respective branching conditions. This is the learned insight that a decision tree provides on the input-space, by telling us that trajectories initiating from this space is likely to violate the specification. Consequently, using the explanation condition, a specified number of random simulations are computed. The last step of the algorithm is to check whether the simulations provide us with the required number of counterexamples. If so, the algorithm terminates by outputting the counterexamples. Otherwise, we append the dataset with the generated simulation traces and retrain and refine the decision tree model again. The steps of generating explanations and computing random simulations are repeated. The proposed algorithm terminates when the desired number of counterexamples are obtained, or when the execution budget is exceeded. We consider the execution budget as the number of retraining iterations permitted for the algorithm. Algorithm \ref{algo:ODT} describes the falsification procedure,  referred to as Decision Tree-based Falsification (\textsc{DTFal}). The left part of Figure \ref{fig:proposed_arch} illustrates the \textsc{DTFal} procedure, the details of which are discussed in the subsequent sections.

\subsection{CPS to Decision-Tree}
\subsubsection{Dataset Generation}
Unlike the earlier dataset, which focused on capturing the system's general input-output behavior, this dataset enables us to classify the trajectories based on their robustness measure. This specialization enhances the decision tree’s ability to identify regions of the input space where the system is likely to violate the safety specification. The dataset is generated from the CPS using the simulators discussed in Section \ref{cpstodataset} together with an STL monitoring tool Breach~\cite{donze2010breach}, which evaluates the robustness of the trajectory with respect to a specified STL safety specification. The input features of the dataset are the components of the initial configuration of the CPS $\mathcal{I}$, that is an initial state $\langle x_i \rangle$, and a piecewise constant input signal $\langle u^i \rangle \in \mathbb{C}(\mathbb{R})$, represented as a vector, element of $\mathbb{R}^{k \times m}$. All trajectories are computed over the same time-horizon and therefore, time-horizon is not captured in the dataset. The output feature is a robustness measure of the trajectory, computed with respect to the given STL safety specification.  The structure of the dataset for building a decision tree is shown in Table \ref{tab:dataset_rob}.

\begin{table}[h]
    \caption{The dataset $\D$ of N trajectories for an STL safety specification $\varphi$. Each row is treated as labeled data for Decision-Tree construction. $\rho(\varphi;\Gamma^i)$ is the robustness value of the $i$-th trajectory.}
    \footnotesize
    \label{tab:dataset_rob}
    \centering
    \begin{tabular}{cccc}
       Initial state & Input signal & Output \\
       \hline
        $\langle x_0^1$ $\rangle$ &  $\langle u_0^1,\dots, u_{k-1}^1\rangle$ & $\rho(\varphi; \Gamma^1)$ \\[5pt]
        $\langle x_0^2\rangle$ & $\langle  u_0^2,\dots, u_{k-1}^2\rangle$ & $\rho(\varphi; \Gamma^2)$ \\[5pt]
        % $\dots$ & $\dots$ & $\dots$ \\
        $\langle x_0^3\rangle$ & $\langle  u_0^3,\dots, u_{k-1}^3\rangle$ & $\rho(\varphi; \Gamma^3)$ \\
        $\dots$ & $\dots$ & $\dots$ \\
        $\dots$ & $\dots$ & $\dots$ \\[5pt]
        $\langle x_0^N\rangle$ & $\langle  u_0^N,\dots, u_{k-1}^N\rangle$ & $ \rho(\varphi; \Gamma^N)$ \\
        % $\dots$ & $\dots$ & $\dots$ \\
        % $\langle x_0^N\rangle$ & $\langle  u_0^N,\dots, u_{k-1}^N\rangle$ & $\langle \rho(\varphi; \Gamma^N,j\Delta t)\rangle$ \\
    \end{tabular}
\end{table}

\subsubsection{Decision-Tree Building}
%describe how we build a decision tree model. describe the details of the CART algorithm.
We used \emph{sklearn} (Scikit-learn), a data modeling library in Python to build the Decision Tree. Specifically, the \emph{DecisionTreeRegressor} method has been used. The method uses the CART~\cite{breiman2017classification} algorithm that builds a binary tree in which leaf nodes represent the predicted output features and intermediate nodes, also known as decision nodes, are where decision choices are made. We used the default parameter settings in \emph{DecisionTreeRegressor}.

When the algorithm meets a stopping criterion in a node, the node is designated as a leaf node. The value of a leaf node is the mean of the datapoint of the target attribute in that node, in our case, which is the robustness value of the trajectories.

\subsection{Generating Explanations to Safety Property Violation}

\subsubsection{Finding Falsifying Leaf Nodes}

Once we approximate the actual CPS with the Decision Tree $\M'$ from the dataset $\D$, our algorithm generates an explanation for the violation of a given STL safety specification by $\M'$. The initial phase for generating the explanation is the retrieval of all leaf nodes from the decision tree $\M'$, followed by filtering the leaf nodes that characterize trajectories that are likely to violate the safety specification. Every leaf node is associated with a predicted robustness measure of the trajectories that map to that node. If a leaf node's robustness measure is less than 0, it classifies the trajectories that are safety violating. We designate such a node as \emph{falsifying leaf node}. 

\begin{definition}[Falsifying leaf node]
    A falsifying leaf node of the decision tree $\M'$ is defined as a leaf node $\mathcal{L}$ such that $\mathcal{L}_{\rho} < 0$, where $\mathcal{L}_{\rho}$ is the node's predicted robustness measure.
\end{definition}

\noindent Consider a decision tree in Figure \ref{fig:Dt-2.97} for an illustration. The only \emph{falsifying leaf node} in this example is highlighted in Red. If no falsifying leaf node is found in the decision tree, our algorithm finds the \emph{nearest falsifying leaf node(s)}, that is, nodes with robustness value closest to 0. The \emph{nearest falsifying leaf nodes} are then referred to generate an explanation. The definition of  \emph{nearest falsifying leaf node} is as follows: 

\begin{definition}[Nearest falsifying leaf node]
    Given a decision tree $\M'$ and given that $\M'$ has no falsifying leaf node, we define a nearest falsifying leaf node $\mathcal{L}_{near}$ to be a leaf node with predicted robustness value closest to zero.
    % $$\mathcal{L}_{near} = \arg\min_{l \in leaf}l_\rho$$
    $$\mathcal{L}_{near} = \arg\min_{l \in leaf}l_\rho$$
    where $l_\rho$ is the predicted robustness value of leaf $l$, and $l_{\rho}$ is its predicted robustness measure. 
\end{definition}
The illustrative decision tree in Figure \ref{fig:DT-3.51} shows the nearest falsifying leaf node in Red. Algorithm \ref{algo:find_nearest_leaf} describes the procedure for finding the nearest falsifying leaf nodes. 

\begin{algorithm}[h]
    \footnotesize
    \caption{\textsc{Find\_Nearest\_Leaf\_Node}{($\mathcal{M}'$)}}
    \begin{algorithmic}
        \Require A Decision Tree $\mathcal{M}'$.
        \Ensure A set of falsifying leaf nodes $\mathcal{L}_{near}$.
        \Initialization $\mathcal{L}_{near} \gets \emptyset$ ;  $m = \infty$ \Comment{Initialize min robustness to infinity.}
        \State $leaf \gets $ \textsc{Find\_All\_Leafs}($\mathcal{M}'$)
        \For{$l \in leaf$}
            \If{$|l_{\rho}| < m$}
                \State $m = |l_{\rho}|$ \Comment{Update minimum robustness value.}
            \EndIf
        \EndFor
        \For{$l \in leaf$}
            \If{$|l_{\rho}| = m$}
                \State $\mathcal{L}_{near} \gets \mathcal{L}_{near} \cup \{l\}$ \Comment{Add leaf nodes with robustness value $m$.}
            \EndIf
        \EndFor
        \State \Return $\mathcal{L}_{near}$
    \end{algorithmic}
    \label{algo:find_nearest_leaf}
\end{algorithm}

\subsubsection{Generating Explanations} Once a \emph{falsifying leaf node} is identified, our algorithm generates an explanation to violating a safety specification. To place the definition of explanation, we present a few notations. The input features of a decision tree are represented by the set $X = \{ x_1, x_2,\ldots, x_n\}$ of $n$ real-valued variables. A valuation $v$ is an assignment of a real value to each variable $x \in X$. We now define an explanation, the central idea to the falsification algorithm. 

\begin{definition}[explanation]
    Given a decision tree $\M'$ with a set $\X$ of $n$ input features and an STL safety specification $\varphi$, an explanation for violating $\varphi$ by $\M'$ is defined as a predicate $Exp_{\M'}(\X)$ over the free variables in $\X$ such that there exists a valuation $v$ of $\X$, where $Exp_{\M'}(v)$ is true and $\M'(v) < 0$. 
\end{definition}

\begin{figure*}[h]
    \centering
    \begin{subfigure}[b]{.45\linewidth}
       \centering
      {\includegraphics[scale=0.45]{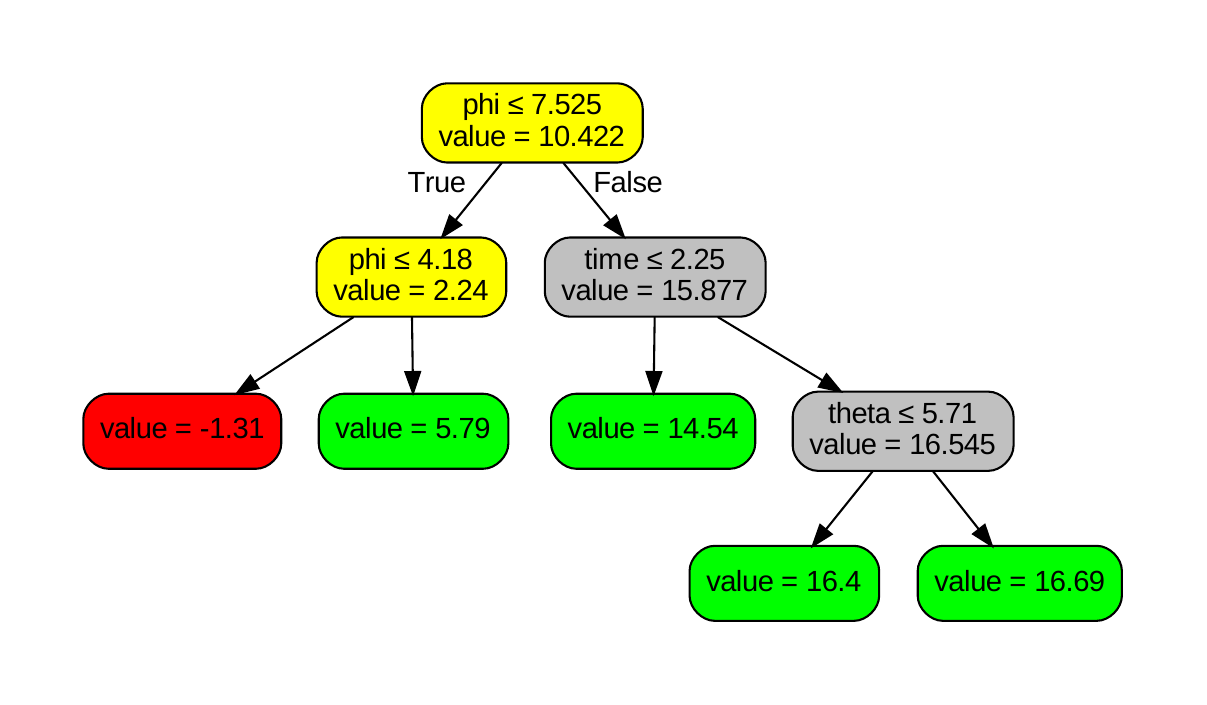}}
      \caption{}
      \label{fig:Dt-2.97}
    \end{subfigure}
    \begin{subfigure}[b]{.49\linewidth}
       \centering
      {\includegraphics[scale=0.45]{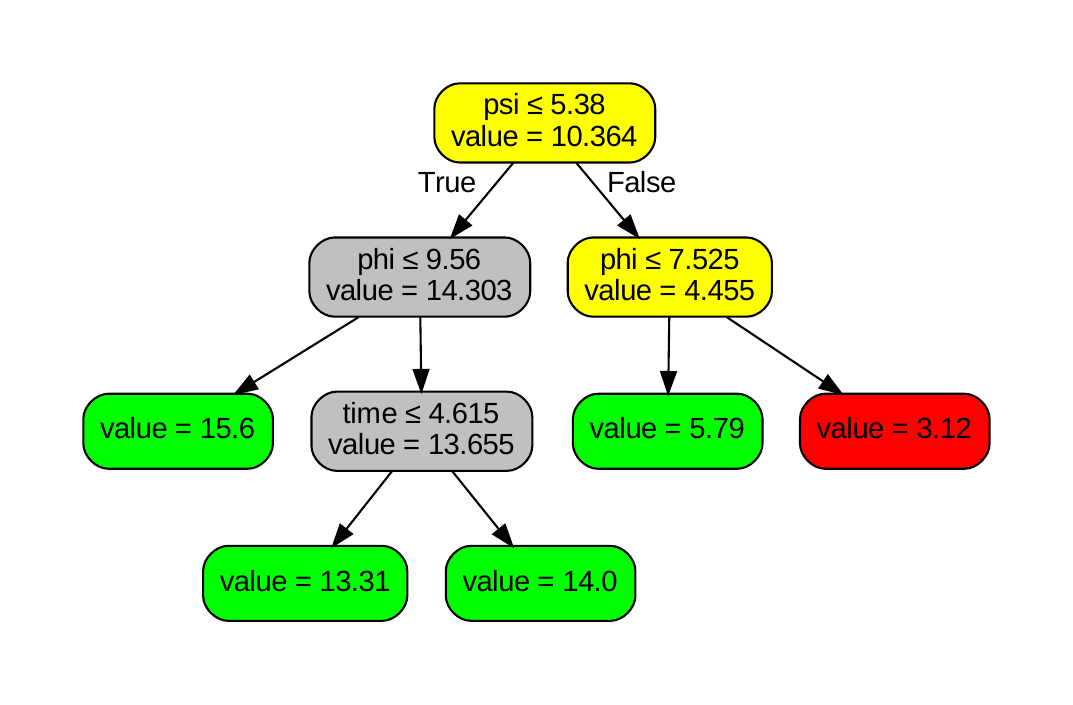}}
      \caption{}
      \label{fig:DT-3.51}
    \end{subfigure}
    \caption{Demonstrating the explanation generation process for a safety property $\square_{[0, 15]} (x > 0)$. a) This figure shows the explanation generation process when a falsifying leaf node is present in the decision tree. The explanation is derived from the path that traces backwards from the falsifying node depicted in Red to the root node. b) When no falsifying leaf node is present in a decision tree, the nearest falsifying leaf node, highlighted in Red is considered to generate the explanation.}
    \label{fig:explanation_demo}
\end{figure*}

\noindent The set of all valuations that satisfy the explanation predicate describes a region of inputs $\subseteq  \mathcal{I}$ to the decision tree which contains a counterexample to the safety specification. Our algorithm constructs an explanation by traversing the decision tree from the \emph{falsifying leaf node} to the root, conjoining the decision-making condition at each intermediate node, to form a predicate that is a conjunction of linear constraints. The process of collecting the decision-making conditions along the path from \emph{falsifying leaf node} to the root is discussed in Algorithm \ref{algo:find_exp}. Initially, the explanation predicate is set to $true$. While traversing upward in the tree from a leaf node, if the current node is \emph{left child}, the decision-making condition of its parent node is conjoined with the explanation predicate. On the other hand, if the current node is a \emph{right child}, the negation of the decision-making condition from its parent node is conjoined with the explanation, since moving to the right branch indicates that the decision condition is not satisfied by the input. Consider the decision tree depicted in Figure \ref{fig:Dt-2.97} and a safety property $\varphi = \square_{[0, 15]} (x > 0)$ for an illustration. The first step of the explanation generation process is to identify the falsifying leaf node among all leaf nodes. In Figure \ref{fig:Dt-2.97}, the Green nodes represent the leaf nodes, while the Red one is the falsifying leaf node with a predicted robustness value $\leq 0$. The next step is to collect the decision-making conditions on the input attributes along the path shown in the figure (starting from the red node followed by two consecutive yellow nodes). Two decision-making conditions ($phi \leq 4.18$ and $phi \leq 7.525$) are obtained which are conjoined because the falsifying leaf node and the intermediate node are the left child of their respective parent nodes. Therefore, the explanation for the falsification of $\varphi$ is ($phi \leq 4.18$ $\land$ $phi \leq 7.525$). On the other hand, when a falsifying leaf node is not present in the decision tree, our algorithm identifies the nearest falsifying leaf nodes from which the explanation is generated. Figure \ref{fig:DT-3.51} shows a decision tree where there in no falsifying leaf node for the safety property $\varphi$. In this figure, the closest falsifying leaf node is shown in Red which is used to generate the explanation. The explanation using this nearest falsifying leaf node is ($phi > 7.525$ $\land$ $psi > 5.38$), the leaf and intermediate node being the right child of their respective parents.

\begin{algorithm}[]
    \footnotesize
    \caption{DTFal Algorithm for Falsification of CPS.}
    \begin{algorithmic}[1]
        \Require CPS model $\M:\P \times \mathbb{C}(\mathbb{R}) \times \mathbb{R} \to \mathbb{R}^n$, an initial set of configurations $\mathcal{I} \subseteq \P \times \mathbb{C} (\mathbb{R}) \times \mathbb{R}$, an STL safety specification $\varphi$, number of initial trajectories $\mathcal{N}$, maximum retrain $epoch$, random simulation R, and minimum counterexample $min\_ce$. 
        \Ensure The desired counterexamples $CE$ if exists, otherwise $Failure$. 
        % \textcolor{red}{Is it always return CE means the model is unsafe}.
        \Initialization $\mathcal{D} \gets \emptyset$, $CE = 0$
        \State $\mathcal{D}$ $\gets$ $\mathcal{D}$ $\cup$ \textsc{Generate\_Training\_Samples}($\mathcal{M}$, $\mathcal{I}$, $\varphi$, $\mathcal{N}$)\label{odt_data_gen}\Comment{Generating dataset.}

        \For{$i = 0$ to $epoch$} \label{odt_main_loop}
            \State $\mathcal{M}'$ $\gets$ \textsc{CART\_Algorithm}($\mathcal{D}$) \label{odt_cart}\Comment{Described in CART algorithm.}
            \State $\mathcal{L}$ $\gets$ \textsc{Find\_Falsifying\_leaf\_node}($\mathcal{M}'$)\label{odt_fal_node}
            % \If{$\mathcal{L}_{u}$ = $\emptyset$} 
            \If{($\mathcal{L} = \emptyset$)} \label{odt_empty_leaf_nodes}
                \State $\mathcal{L} \gets$ \textsc{Find\_Nearest\_leaf\_node}($\mathcal{M}'$)\label{odt_near_fal_node}
            % \EndIf
            \EndIf
           % \State \textcolor{red}{$\mathcal{I} = \emptyset$}
            \For{leaf in $\mathcal{L}$} \label{odt_leaf_node_process}
                \State $Exp_{\mathcal{M}'} \gets$ \textsc{Gen\_Explanations}($\mathcal{M}'$, leaf) \label{tr_to_root}\Comment{Generating explanations }
                %\State $\mathcal{I}$ $\gets$ $\mathcal{I}$ $\cap$ $\mathcal{I}'$\Comment{Making $\mathcal{I}$ more precise.}
                \State $ce, traj\_data = $ \textsc{Simulator}($\mathcal{M}$, $Exp_{\mathcal{M}'}$, $\varphi$, R) \label{ran_simu}\Comment{Searching counterexamples by R simulations.}
                \State $CE$ $\gets$ $CE$ $+$ $ce$ 
                \If{ ($CE \geq min\_ce$)}\label{odt_req_ce}
                    \State \Return $CE$ 
                % \Comment{\textcolor{red}{where the counter increased?}}
                \EndIf
                % \If{$\mathcal{CE}$ count increases}
                    
                    % \Comment{\textcolor{red}{What if the condition is not satisfied?}}
                % \EndIf
                \State $\mathcal{D}$ $\gets$ $\mathcal{D}$ $\cup$ $traj\_data$
            \EndFor \label{odt_leaf_nodes_process_end}
        \EndFor
        \State \Return $Failure$
    \end{algorithmic}
    \label{algo:ODT}
\end{algorithm}

\begin{algorithm}
    \footnotesize
    \caption{\textsc{Gen\_Explanations}{($\mathcal{M}'$, $leaf$)}}
    \begin{algorithmic}[1]
        \Require A Decision Tree $\mathcal{M}'$, and a falsifying leaf node $leaf$.
        \Ensure An explanation $Exp_{\mathcal{M}'}$ from $\mathcal{M}'$ for a falsifying leaf node $leaf$.
        \Initialization $Exp_{\mathcal{M}'} \gets True$
        % \State $Exp_{\mathcal{M}'} \gets True$
        \State $current\_node = leaf$ 
        \While{($current\_node \neq Null$)}\Comment{Iterate until the current node becomes root.}
            \If{($current\_node.leftChild$)}\Comment{If current node is the left child of its parent.}
                \State $current\_node = current\_node.parent$
                \State $Exp_{\mathcal{M}'} \gets Exp_{\mathcal{M}'}  \wedge current\_node.condition$
            \Else \Comment{If current node is the right child of its parent.}
                \State $current\_node = current\_node.parent$
                \State $Exp_{\mathcal{M}'} \gets Exp_{\mathcal{M}'}  \wedge \neg (current\_node.condition)$
            \EndIf
        \EndWhile
        \State \Return $Exp_{\mathcal{M}'}$
    \end{algorithmic}
    \label{algo:find_exp}
\end{algorithm}

\subsection{Generating Counterexamples}

\noindent Algorithm \ref{algo:ODT} illustrates the \textsc{DTFal} procedure, which takes the CPS model $\M$, an initial condition $\mathcal{I}$, a STL safety specification $\varphi$, the number of initial trajectories $\mathcal{N}$, execution budget $epoch$, and required counterexample $min\_ce$ as inputs. The function \textsc{Generate\_Training\_Samples}($\mathcal{M}$, $\mathcal{I}$, $\varphi$, $\mathcal{N}$) in line \ref{odt_data_gen} generates the dataset from the initial condition $\mathcal{I}$ for $\mathcal{N}$ trajectories and stores into $\mathcal{D}$. The for loop in line \ref{odt_main_loop} iterates over the execution budget $epoch$. The decision tree $\mathcal{M}'$ is constructed from the dataset $\mathcal{D}$ in line \ref{odt_cart} using the CART algorithm. Line \ref{odt_fal_node} finds the falsifying leaf nodes ($\mathcal{L}$). Line \ref{odt_near_fal_node} states that if $\M'$ has no falsifying leaf node, then our algorithm finds the nearest falsifying leaf nodes. The for loop in lines \ref{odt_leaf_node_process} - \ref{odt_leaf_nodes_process_end} iterates over the leaf nodes $\mathcal{L}$, identifying the explanation ($Exp_{\mathcal{M}'}$) to the violation of $\varphi$ and subsequently determining the true counterexamples. Line \ref{tr_to_root} finds the explanation ($Exp_{\mathcal{M}'}$) using the procedure defined in Algorithm \ref{algo:find_exp}. In the procedure \textsc{Simulator} in line \ref{ran_simu}, $R$ many valuations are randomly selected that satisfy the explanation $Exp_{\mathcal{M}'}$. From these R valuations, R random trajectories are simulated, and their robustness is observed to see how many qualify to be counterexamples. This function also generates the dataset ($traj\_data$) from the R trajectories, which is used to retrain the decision tree, if required. If the number of counterexamples ($CE$) reaches the required minimum threshold within the execution budget, the algorithm terminates and displays the $CE$ counterexamples. Otherwise, failure is returned.

\section{Results}

\begin{table*}[h]
    
    \caption{Safety specifications are given in STL. Dims is the sum of the number of system variables and input variables in the CPS. DoD is the percentage of safety-violating trajectories of $2000$ simulated trajectories.}
    %\footnotesize
    \centering
    \begin{adjustbox}{width=.8\textwidth}
    \begin{tabular}{|c|c|c|c|c|}
        \hline
        Instance & CPS & Dims & DoD & Safety Specification in Signal Temporal Logic\\
        \hline
        TT1 & \multirow{4}{*}{Two tanks}  & 3 & $0.32\%$ & $\square_{[0, 10]}\neg\bigl ((x1 \geq 0)\land (x1 \leq 0.40) \land (x2 \geq -0.500)\land (x2 \leq -0.465)\bigl )$ \\
        \cline{1-1} \cline{3-5}
        TT2 & & 3 & $0.83\%$ &  $\square_{[0, 10]}\neg\bigl ((x1 \geq -0.20) \land (x1 \leq 0.20) \land (x2 \geq 0.31) \land (x2 \leq 0.35)\bigl )$ \\
        \cline{1-1}\cline{3-5}
        TT3 & & 3 & $2.68\%$ & $\square_{[0, 10]}\neg\bigl ((x1 \geq -0.20) \land (x1 \leq 0.20) \land (x2 \geq 0.30) \land (x2 \leq 0.35)\bigl )$ \\
         \cline{1-1}\cline{3-5}
       TT4 & & 3 & $2.95\%$ & $\square_{[0, 10]}\neg\bigl ((x1 \geq 1) \land (x1 \leq 1.5) \land (x2 \geq -0.4) \land (x2 \leq -0.23)\bigl )$ \\
         \hline
        NAV1 & \multirow{3}{*}{Navigation} & 3 & $0.003\%$ & $\square_{[0, 50]}\neg\bigl ((x1 \geq 7) \land (x1 \leq 8) \land (x2 \geq 9) \land (x2 \leq 10)\bigl )$ \\
         \cline{1-1}\cline{3-5}
        NAV2 & & 3 & $0.07\%$ & $\square_{[0, 50]}\neg\bigl ((x1 \geq 22) \land (x1  \leq 23) \land (x2 \geq 11) \land (x2 \leq 12)\bigl )$ \\
         \cline{1-1}\cline{3-5}
        NAV3 & & 3 & $2.77\%$ & $\square_{[0, 50]}\neg\bigl ((x1 \geq 11) \land (x1 \leq 12) \land (x2 \geq 16) \land (x2 \leq 17)\bigl )$ \\
         \cline{1-1}\cline{3-5}
         \hline
        OSC1 & \multirow{3}{*}{Oscillator} & 3 & $0.15\%$ & $\square_{[0, 10]}\neg\bigl ((p \geq 0) \land (p \leq 0.1) \land (q \geq 0.13485) \land (q \leq 0.15)\bigl )$ \\
         \cline{1-1}\cline{3-5}
        OSC2 & & 3 & $0.29\%$ & $\square_{[0, 10]}\neg\bigl ((p \geq -0.50) \land (p \leq -0.45029) \land (q \geq 0.1) \land (q \leq 0.1968)\bigl )$ \\
         \cline{1-1}\cline{3-5}
        OSC3 & & 3 & $0.41\%$ & $\square_{[0, 10]}\neg\bigl ((p \geq 0.100) \land (p \leq 0.193) \land (q \geq -0.30) \land (q \leq -0.25)\bigl )$ \\
         \cline{1-1}\cline{3-5}
         \hline
         BB1 & Bouncing ball & 3 & $1.2\%$ & $\square_{[0, 10]} \neg((v \geq -1) \land (v \leq 1) \land (x \geq 1) \land (x \leq 2))$ \\
         \hline
         ACC1 & ACC & 10 & $0\%$ & $\bigwedge_{i=0..3}\square_{[0, 10]} (x_i > x_{i+1})$ \\
         \hline
         F16 & F16 & 16 & $0.1\%$ &  $\square_{[0, 15]} (altitude > 0)$ \\
         \hline
         CC1 & \multirow{6}{*}{CC} & 7 & $0.60\%$ &  $\square_{[0, 100]} (y_5 - y_4 \leq 40)$ \\
         \cline{1-1}\cline{3-5}
         CC2 &  & 7 & $0\%$ &  $\square_{[0, 70]} \Diamond_{[0,30]} y_5 - y_4 \geq 15$ \\
         \cline{1-1}\cline{3-5}
         CC3 &  & 7 & $1.15\%$ & $\square_{[0, 80]} ((\square_{[0,20]}y_2 - y_1 \leq 20) \lor (\Diamond_{[0,20]} y_5 - y_4 \geq 40))$ \\
         \cline{1-1}\cline{3-5}
         CC4 &  & 7 & $0\%$ & $\square_{[0, 65]} \Diamond_{[0,30]} \square_{[0,5]} y_5 - y_4 \geq 8$ \\
         \cline{1-1}\cline{3-5}
         CC5 &  & 7 & $1.45\%$ & $\square_{[0, 72]} \Diamond_{[0,8]} ((\square_{[0,5]} y_2 - y_1 \geq 9) \to (\square_{[5, 20]} y_5 - y_4 \geq 9))$ \\
         \cline{1-1}\cline{3-5}
         CCx &  & 7 & $0.55\%$ &  $\bigwedge_{i=1..4} \square_{[0, 50]} (y_{i+1} - y_{i} > 7.5)$ \\
         \hline
         AT1 & \multirow{10}{*}{AT} & 5 & $0\%$ & $\square_{[0, 20]} (v < 120)$ \\
        % \cline{2-3}
         %& AT-P2                                          & $\square_{[0, 10]}( \omega < 4750)$ \\
          \cline{1-1}\cline{3-5}
        AT2 & & 5 & $4.55\%$ & $\square_{[0, 10]} (\omega < 4750)$ \\
         \cline{1-1}\cline{3-5}
        AT51 & & 5 & $9.05\%$ & $\square_{[0, 30]} ((\lnot g1 \land \circ~g1) \to \circ~\square_{[0, 2.5]} g1)$, \\
         & & & & where [$\circ~\phi \equiv \diamond_{[0.001, 0.1]}~\phi$] \\
         \cline{1-1}\cline{3-5}
        AT52 & & 5 & $5.55\%$ & $\square_{[0, 30]} ((\lnot g2 \land \circ~g2) \to \circ~\square_{[0, 2.5]} g2)$ \\
         \cline{1-1}\cline{3-5}
        AT53 & & 5 & $6.05\%$ & $\square_{[0, 30]} ((\lnot g3 \land \circ~g3) \to \circ~\square_{[0, 2.5]} g3)$ \\
         \cline{1-1}\cline{3-5}
        AT54 & & 5 & $3.1\%$ & $\square_{[0, 30]} ((\lnot g4 \land \circ~g4) \to \circ~\square_{[0, 2.5]} g4)$ \\
         %\hline
         \cline{1-1}\cline{3-5}
         AT6a & & 5 & 0.25\% & $(\square_{[0, 30]} \omega < 3000) \rightarrow (\square_{[0, 4]} v < 35)$ \\
         \cline{1-1}\cline{3-5}
         AT6b & & 5 & 0.1\% & $(\square_{[0, 30]} \omega < 3000) \rightarrow (\square_{[0, 8]} v < 50)$ \\
         \cline{1-1}\cline{3-5}
         AT6c & & 5 & $0.15\%$ & $(\square_{[0, 30]} \omega < 3000) \rightarrow (\square_{[0, 20]} v < 65)$ \\
          \cline{1-1}\cline{3-5}
         AT6abc & & 5 & $0.35\%$ & AT6a $\land$ AT6b $\land$ AT6c \\
         \hline
        % SC & SC & 5 & $ 0\%$ & $\square_{[30, 35]} (87 \leq pressure \land pressure \leq 87.50)$ \\
        %  \hline
         \multirow{3}{*}{AFC27} & \multirow{5}{*}{AFC} & 4 & $1.5\%$ & $ \square_{[11, 50]} (( rise \lor fall) \rightarrow (\square_{[1, 5]} |\mu| < 0.008))$\\
         & & & & where $rise = (\theta < 8.8 \wedge \Diamond_{[0, 0.05]} \theta > 40.0)$ \\
         & & & & $fall = (\theta > 40.0 \land \Diamond_{[0, 0.05]} \theta < 8.8)$ \\
         \cline{1-1}\cline{3-5}
        AFC29 &  & 4 & $ 7.95\%$ & $\square_{[11, 50]} (\mu < 0.007)$ \\
         \cline{1-1}\cline{3-5}
        AFC33 &  & 4 & $ 8.2\%$ & $\square_{[11, 50]} (\mu < 0.007)$ \\
         \hline
         SB1 & SB-1 & 3 & $0 \%$ & $\square_{[0, 24]}$ $ ( b < 20)$   \\
         \hline
         SB2 & SB-2 & 2 & $0\%$ & $\square_{[0, 18]} (b > 90 \lor \Diamond_{[0, 6]} b < 50)$  \\
         \hline
         SB3 & SB-3 & 3 & $0.05\%$ & $(\Diamond_{[6, 12]} b > 10) \rightarrow (\square_{[18, 24]} b > -10)$  \\
         \hline
         SB4 & SB-4 & 4 & $0\%$ & $\square_{[0, 19]} ((\square_{[0, 5]} b_1 \leq 20) \lor (\Diamond_{[0, 5]} b_2 \geq 40))$  \\
         \hline
         SB5 & SB-5 & 3 & $0\%$ & $\square_{[0, 17]} (\Diamond_{[0, 2]} \neg((\square_{[0, 1]} b_1 \geq 9) \lor (\square_{[1, 5]} b_2 \geq 9)))$  \\
         \hline
    \end{tabular}
    \end{adjustbox}
    \label{tab:safetySpecification}
\end{table*}

\noindent \textbf{Benchmarks} The framework has been evaluated on falsification problem instances from 9 CPS out of which 6 have linear dynamics (Navigation~\cite{FehnkerI04}, Oscillator~\cite{FLGDCRLRGDM11}, Two tanks~\cite{DBLP:conf/hicss/Hiskens01}, Bouncing ball, Adaptive Cruise Controller (ACC)~\cite{DBLP:conf/arch/BuAAMRWZ20}, and Chasing Cars (CC)~\cite{10.5555/646880.710449}) and three have non-linear dynamics (Aircraft Ground Collision Avoidance System (F16)~\cite{DBLP:conf/adhs/HeidlaufCBB18}, Automatic Transmission (AT)~\cite{DBLP:conf/cpsweek/HoxhaAF14}, and Fuel Control of an Automotive Powertrain (AFC)~\cite{jin2014powertrain}. 
Recently, a procedure for creating synthetic benchmarks has been recently introduced in \cite{yan2025automated} for testing the strength of falsification methods. We considered five Synthetic Benchmarks (SB) from that paper, each involving an LSTM model trained on a synthetic dataset designed for a specific safety property. We now briefly describe the models and the specifications, some of which are from the annual friendly tool competition ARCH-COMP 2024~\cite{khandait2024arch}, in the falsification category. The safety specifications in signal temporal logic (STL) are shown in Table \ref{tab:safetySpecification}.

\subsection{Implementation and Experimental Setup}

% \paragraph{Setup for Evaluating Falsification Instances}
The framework is implemented in a tool \textsc{FlexiFal} and offers the two falsification strategies to a user, namely \textsc{NNFal} and \textsc{DTFal}. The tool provides a command-line interface. The source code of \textsc{FlexiFal}, the command-line options and user instructions are made publicly available on GitLab [\url{https://gitlab.com/Atanukundu/FlexiFal}]. We evaluate \textsc{FlexiFal} on Ubuntu 22.04.5, 64 GB RAM with 6 GB GPU, and a 5 GHz 12-core Intel i7-12700K processor. For the evaluation of \textsc{NNFal}, we build the neural network surrogates of CPS in advance. The neural networks were trained in Google Colab, with Intel Xeon 2.20 GHz CPU, 12.7 GB RAM with 15 GB Tesla T4 GPU, and Ubuntu 20.04. In order to quantify the difficulty level of a falsification instance, specifically for the hybrid automata instances, we define a degree of difficulty (DoD) metric, which is the percentage of the falsifying trajectories in $N$ randomly simulated trajectories. The lower the DoD, the harder it is to falsify the instance. All experiments have 1 hour timeout and a maximum memory usage limit of 6 GB. We test each falsification instance ten times and report the success rate and the average falsification time of the successful tests. In the following performance comparison table, the gray cells highlight the algorithms or tools that demonstrate the best time to falsify the respective instances. The results of the experiments provide a response to the following research questions. 

\noindent \textbf{RQ1:} Which out of \textsc{DTFal} and \textsc{NNFal} is more effective in falsifying CPS instances?

\noindent \textbf{RQ2:} How does \textsc{FlexiFal} compare with state-of-the-art CPS falsification tools?

\noindent \textbf{RQ3:} Can \textsc{FlexiFal} find multiple counterexamples efficiently?

\subsection{\textbf{RQ1}: \textsc{NNFal} vs \textsc{DTFal}}

\begin{table*}[htbp]
    \caption{FR stands for falsification rate, the number of successful runs out of 10 for which our framework finds a counterexample (CE). In \textsc{DTFal}, DG denotes the time taken to generate the training dataset, while Train represents the time to train the decision tree, including the retrainings if necessary. Search refers to the time  to find counterexamples using the generated explanations. Fal represents the total time to identify a counterexample, calculated as the sum of DG, Train, and Search time. In \textsc{NNFal}, TT is the neural network training time. MR shows the maximum number of search refinements to eliminate spurious CEs before finding a valid one. Avg time is the average time that \emph{pgd} takes to find a valid CE, calculated over successful runs. Val is the average time taken by \textsc{NNFal} to validate the CE on the actual CPS across the successful runs, incorporating the time required for spurious ce checking and the specification modifications before the true CE is encountered. Fal Time in an adversarial attack algorithm is the sum of Avg and Val over the successful runs. Fal Time in the DNN verification tool is similarly the sum of falsification time and validation time for the respective instances. Times are reported in seconds except in the TT column, which shows time in hours. NA indicates that the respective algorithm does not support the instance.}
    \centering
    \begin{adjustbox}{width=1\textwidth}
    \begin{tabular}{|c|c|c|c|c|c|c|c|c|c|c|c|c|c|c|c|c|}
        \hline
        \multirow{4}{*}{Instance} & \multicolumn{5}{c|}{\textsc{DTFal}} & \multicolumn{11}{c|}{\textsc{NNFal}} \\
        \cline{2-17}
        & \multirow{2}{*}{FR} & \multicolumn{4}{c|} {Time}  & \multirow{2}{*}{TT} & \multicolumn{5}{c|}{Adversarial Attack Algorithm (\emph{pgd})} & \multicolumn{5}{c|}{DNN Verification Tool (\emph{nnenum})}\\
        \cline{3-6}\cline{8-17}
        & & DG & Train & Search & Fal & & FR & MR  & Avg Time & Val & Fal Time & FR & MR & Time & Val & Fal Time\\
        \hline
        TT1 & 10 & 6.6 & 0.35 & 0.17 & 7.12 & \multirow{4}{*}{4.3} & 10 & 1 & 150.80 & 0.32 & 151.12 & 10 & 0 & 1.78 & 0.26 & \cellcolor[gray]{0.8} 2.04 \\
        \cline{1-6}\cline{8-17}
        TT2 & 10 & 1.64 & 0.03 & 0.18 & \cellcolor[gray]{0.8}1.85 & & 10 &  0 & 996.66 & 0.25 & 996.91 & 10 & 0 & 4.00 & 0.26 & 4.26 \\
        \cline{1-6}\cline{8-17}
        TT3 & 10 & 1.67 & 0.03 & 0.17 & \cellcolor[gray]{0.8}1.87 & & 10  & 0 & 62.94 & 0.25 & 63.19 & 10 & 0 & 3.33 & 0.27 & 3.60 \\
        \cline{1-6}\cline{8-17}
        TT4 & 10 & 1.61 & 0.03 & 0.17 & 1.81 &  & 10  &  0 & 23.41 & 0.28 & 23.69 & 10 & 0 & 1.51 & 0.27 & \cellcolor[gray]{0.8}1.78 \\
         \hline
        Osc1 & 10 & 6.99 & 0.37 & 0.26 & 7.62 & \multirow{3}{*}{0.3} & 10  &  0 & 0.24 &   0.28 & \cellcolor[gray]{0.8} 0.52 & 10 & 0 & 1.60 & 0.31 & 1.91 \\
        \cline{1-6}\cline{8-17}
        Osc2 & 10 & 7.11 & 0.35 & 0.20 & \cellcolor[gray]{0.8} 7.66 & &  &   &  &   & OOM &  &  &  &  & Timeout \\
        \cline{1-6}\cline{8-17}
        Osc3 & 10 & 1.09 & 0.005 & 0.19 & \cellcolor[gray]{0.8} 1.28 & &  10 & 5 & 70.04 &  1.68 & 71.72 & 6 & 9 & 22.47 & 1.85 & 24.32\\
        \hline
        %Two tanks\_NN-2 & P1 & 4& 3 & 2.95 $\%$ & 10 &  0 & 68.73 & 0.29 & 69.02 \\
       %\hline
        NAV1 & 10 & 133.33 & 12.95 & 1.83 & \cellcolor[gray]{0.8}148.11 & \multirow{3}{*}{3.6} & 5 & 3 & 1091.07 & 5.25 & 1096.32 & &  &  &  & Timeout \\ %P1 is set-4
        %\textbf{Simulink models} & & & & & & & & & & & & & & & \\
        \cline{1-6}\cline{8-17}
        NAV2 & 10 & 14 & 0.75 & 1.80 & \cellcolor[gray]{0.8}16.55 & & 7 & 2 & 14.16 &  3.11 &  17.27 &  &  &  &  & Timeout\\ %P1 is set-4
        \cline{1-6}\cline{8-17}
        NAV3 & 10 & 2.88 & 0.04 & 0.86 & $\cellcolor[gray]{0.8}$3.78 & & 10 & 0 & 3.20 & 1.38 & 4.58 & 10 & 3 & 3.65 & 2.68 & 6.33\\
        \hline
         BB1 & 10 & 1.61 & 0.73 & 0.17 & \cellcolor[gray]{0.8} 2.51 & & 0 &  &  &  & Timeout &  &  &  &  & Error\\
        \hline
        ACC1 & 10 & 1.24 & 0.007 & 0.22 & 1.47 & 0.5 & 10 & 3 & 0.09 & 0.40 & \cellcolor[gray]{0.8} 0.49 & 10 & 1 & 1.40 & 0.37 & 1.77 \\
        \hline
        F16 & 10 & 469.52 & 0.176 & 67.40 & \cellcolor[gray]{0.8} 537.09 & 1.4 &  &  &  &  & Timeout &  &  &  &  & OOM \\
        \hline
        CC1 & 10 & 72.33 & 15.48 & 7.53 & 95.34 & \multirow{6}{*}{0.7} & 10 & 0 & 0.11 & 23.54 &  \cellcolor[gray]{0.8}23.65 & 10 & 0 & 2.21 & 23.36 & 25.57 \\
        \cline{1-6}\cline{8-17}
        CC2 & 10 & 97.96 & 11.41 & 17.96 & 127.33 &  &  &  &  &  &  NA &  &  &  &  & NA \\
        \cline{1-6}\cline{8-17}
        CC3 & 10 & 72.88 & 8.88 & 1.27 & \cellcolor[gray]{0.8}83.03 &  &  &  &  &  &  NA &  &  &  &  & NA \\
        \cline{1-6}\cline{8-17}
        CC4 &  &  &  &  & Timeout &  &  &  &  &  &  NA &  &  &  &  & NA \\
        \cline{1-6}\cline{8-17}
        CC5 & 10 & 75.47 & 3.90 & 4.11 & \cellcolor[gray]{0.8}83.48 &  &  &  &  &  &  NA &  &  &  &  & NA \\
        \cline{1-6}\cline{8-17}
        CCx & 10 & 147.72 & 8.98 & 56.34 & 213.04 &  & 10 & 0 & 0.15 & 46.35 & 46.50 & 5 & 0 & 1.70 & 40.70 & \cellcolor[gray]{0.8}42.40 \\
        \hline
        AT1 &  & & & & Timeout & \multirow{10}{*}{4.0} & 1 & 2 & 0.10 & 72.44 &  \cellcolor[gray]{0.8}72.54 & 1 & 4 & 96.79 & 46.45 & 143.24 \\
        \cline{1-6}\cline{8-17}
        AT2 & 10 & 13.95 & 0.30 & 1.90 &  \cellcolor[gray]{0.8}16.15 & &  &  &  &  & OOM &  &  &  &  & Error \\
        \cline{1-6}\cline{8-17}
        AT51 & 10 & 17.74 & 0.22 & 1.15 &  \cellcolor[gray]{0.8}19.11 & &  &  &  &  & NA &  &  &  &  & NA \\
        \cline{1-6}\cline{8-17}
        AT52 & 10 & 14.19 & 0.17 & 2.34 &  \cellcolor[gray]{0.8}16.70 & &  &  &  &  & NA &  &  &  &  & NA \\
        \cline{1-6}\cline{8-17}
        AT53 & 10 & 14.34 & 0.22 & 2.34 &  \cellcolor[gray]{0.8}16.90 & &  &  &  &  & NA &  &  &  &  & NA \\
        \cline{1-6}\cline{8-17}
        AT54 & 10 & 14.19 & 0.14 & 2.04 &  \cellcolor[gray]{0.8}16.37 & &  &  &  &  & NA &  &  &  &  & NA \\
        \cline{1-6}\cline{8-17}
         AT6a & 10 & 42.52 & 2.25 & 0.38 &  \cellcolor[gray]{0.8}45.15 & &  &  &  &  & OOM &  &  &  &  & Timeout \\
        \cline{1-6}\cline{8-17}
        AT6b & 10 & 65.68 & 3.35 & 3.45 &  \cellcolor[gray]{0.8}72.48 & &  &  &  &  & OOM &  &  &  &  & Timeout \\
        \cline{1-6}\cline{8-17}
        AT6c & 10 & 43.19 & 1.34 & 8.20 & 52.73 & & 2 & 1 & 0.01 & 46.23 & \cellcolor[gray]{0.8} 46.24  &  &  &  &  & Timeout \\
        \cline{1-6}\cline{8-17}
        AT6abc & 10 & 43.76 & 1.31 & 1.49 & \cellcolor[gray]{0.8}46.56 & &  &  &  &  &  OOM &  &  &  &  & OOM \\
        \hline
        AFC27 & 10 & 26.85 & 1.33 & 4.84 & \cellcolor[gray]{0.8} 33.02 & \multirow{3}{*}{2.3} &  &  &  &  & NA &  &  &  &  & NA \\
        \cline{1-6}\cline{8-17}
        AFC29 & 10 & 26.95 & 0.88 & 1.89 & \cellcolor[gray]{0.8} 29.72 & &  &  &  &  & Timeout &  &  &  &  & Timeout \\
        \cline{1-6}\cline{8-17}
        AFC33 & 10 & 25.28 & 0.67 & 1.86 & \cellcolor[gray]{0.8} 25.28 & &  &  &  &  & Timeout &  &  &  &  & Timeout \\
        \hline
        SB1 & 5 & 3182.71 & 10.43 & 17.67 & \cellcolor[gray]{0.8}3210.81 & 0.1 &  &  &  &  & OOM &  &  &  &  & Timeout \\
        \hline
        SB2 &  &  &  &  & Timeout & &  &  &  &  & NA &  &  &  &  & NA \\
        \hline
        SB3 & 10  & 52.72 & 0.02 & 11.92 & \cellcolor[gray]{0.8}64.66 & &  &  &  &  & NA &  &  &  &  & NA \\
        \hline
        SB4 & 10 & 1951.45  & 3.14 & 88.97 & \cellcolor[gray]{0.8}2043.56 & &  &  &  &  & NA &  &  &  &  & NA \\
        \hline
        SB5 &  &  &  &  & Timeout & &  &  &  &  & NA &  &  &  &  & NA \\
        \hline
    \end{tabular}
    \end{adjustbox}
    \label{tab:easy_hard_instance}
\end{table*}

\noindent Table \ref{tab:easy_hard_instance} demonstrates the empirical results of executing \textsc{NNFal} and \textsc{DTFal} on several falsification instances. We describe the choice of parameters and a performance comparison.

\paragraph{Evaluation using \textsc{NNFal} } Recall that \textsc{NNFal} provides a repertoire of neural network adversarial attack algorithms and network verification tools via the DNNF and DNNV interface. We report the performance of \textsc{NNFal} using adversarial attack algorithms as well as using neural network verification tools supporting falsification. The falsification instances are evaluated using four attack algorithms, namely \emph{pgd}~\cite{madry2017towards}, \emph{fgsm}~\cite{goodfellow2014explaining}, \emph{bim}~\cite{DBLP:conf/iclr/KurakinGB17a}, and \emph{ddnattack}~\cite{rony2019decoupling}. In the table for \textsc{NNFal}, we show the result of the fastest algorithm to falsify. The fastest in our experiments turns out to be \emph{pgd}. Furthermore, all falsification instances are also evaluated using three neural network verification tools, namely \emph{Reluplex}~\cite{katz2017reluplex}, \emph{Neurify}~\cite{wang2018efficient}, and \emph{Nnenum}~\cite{bak2020improved}. We observe that \emph{Nnenum} is the fastest to falsify in our experiments, and thus we report results on \emph{Nnenum} for \textsc{NNFal}. Our findings indicate that falsification with \textsc{NNFal} generates novel counterexamples that are unseen in the dataset. The results show that \emph{Nnenum} finds counterexamples faster compared to \emph{pgd} in most instances. However, for hard instances (very low DoD), \emph{Nnenum} fails in the falsification task, where \emph{pgd} is successful in a reasonable time. The observations, therefore, do not conclude that verification tools of deep neural networks are more effective than local robustness property falsifiers or vice versa for CPS falsification. Rather, we conclude that they have complementary strengths, and we can therefore reap the benefits of these techniques with our framework. Note that the time to build the neural network is not included in the falsification time for \textsc{NNFal}. We choose to do so because the falsification algorithms need not construct the network every time a new specification is given for falsification. The networks, once constructed, can be reused for falsification tasks. Since \textsc{NNFal} is limited to falsification of reachability specifications, it could not falsify many of the instances in our set which had generic STL properties. Such instances are marked with NA against \textsc{NNFal} in the table. Overall, it could falsify $14$ out of $37$ instances.

\begin{figure*}[htbp]
    \centering
    \caption{Counterexamples generated by \textsc{FlexiFal} using \textsc{NNFal} algorithm.}
    \begin{subfigure}[b]{.3\linewidth}
      \centering
	  {\includegraphics[width=\textwidth]{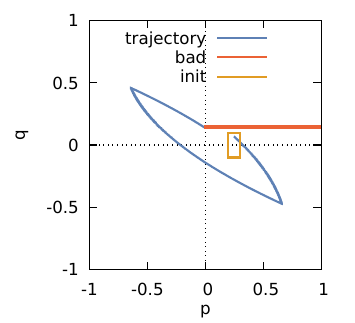}}
      \caption{A trajectory that results in a violation of OSC3 by entering an unsafe region.}
    \end{subfigure}
    \begin{subfigure}[b]{.3\linewidth}
       \centering
      {\includegraphics[width=\textwidth]{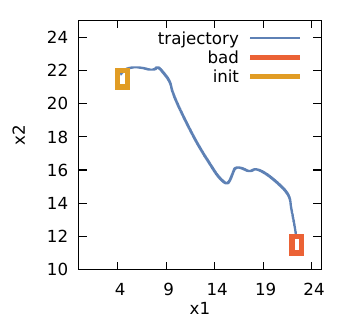}}
      \caption{A navigation trajectory leading to an unsafe region, violating NAV2.}
    \end{subfigure}
    \begin{subfigure}[b]{.3\linewidth}
       \centering
      {\includegraphics[width=\textwidth]{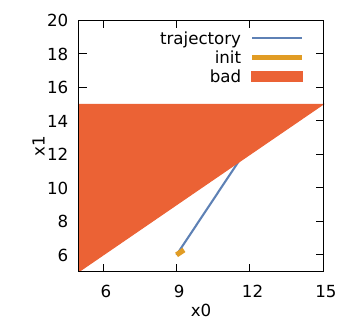}}
      \caption{A positions trajectory of two cars crashing (x0 = x1), violating ACC1.}
    \end{subfigure}
    \label{fig:Traj_NNFal}
% \end{figure*}
% \begin{figure*}[h]
    \centering
    \caption{Counterexamples generated by \textsc{FlexiFal} using \textsc{DTFal} algorithm.}
    \begin{subfigure}[b]{.3\linewidth}
      \centering
	  {\includegraphics[width=\textwidth]{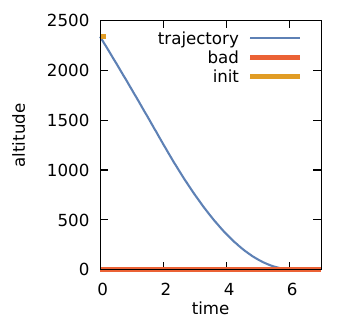}}
      \caption{A flight trajectory of an aircraft crashing at the surface, violating F16.}
    \end{subfigure}
    \begin{subfigure}[b]{.3\linewidth}
       \centering
      {\includegraphics[width=\textwidth]{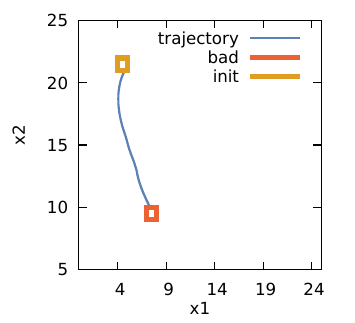}}
      \caption{A navigation trajectory leading to an unsafe region, violating NAV1.}
    \end{subfigure}
    \begin{subfigure}[b]{.3\linewidth}
       \centering
      {\includegraphics[width=\textwidth]{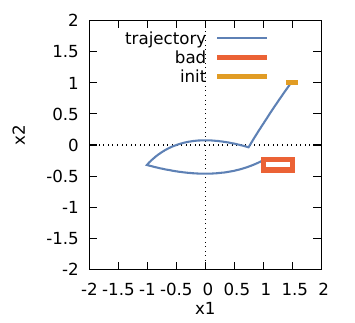}}
      \caption{A trajectory of instance TT4 that intersects with an unsafe region.}
    \end{subfigure}
    \label{fig:Traj_ODFal1}
% \end{figure*}
% \begin{figure*}[h]
    \centering
    \caption{Counterexamples generated by \textsc{FlexiFal} using \textsc{DTFal} algorithm.}
    \begin{subfigure}[b]{.3\linewidth}
      \centering
	  {\includegraphics[width=\textwidth]{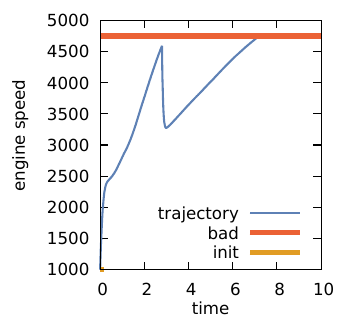}}
      \caption{A trajectory showing the violation of engine speed of the instance AT2.}
    \end{subfigure}
    \begin{subfigure}[b]{.3\linewidth}
       \centering
        {\includegraphics[width=\textwidth]{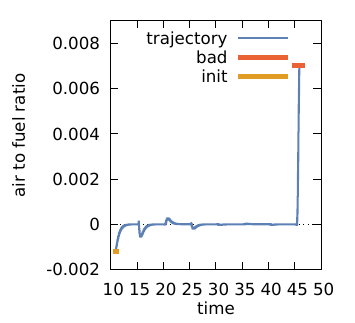}}
        \caption{Air to fuel ratio is leading to the unsafe threshold, violating the instance AFC29.}
    \end{subfigure}
    \begin{subfigure}[b]{.3\linewidth}
       \centering
      {\includegraphics[width=\textwidth]{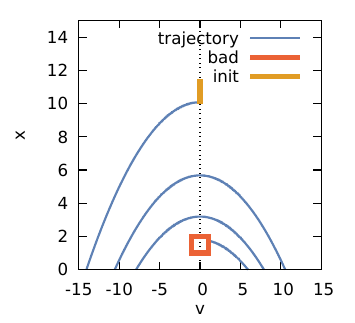}}
      \caption{A position vs. velocity plot of BB1 that results in a safety violation.}
    \end{subfigure}
    \label{fig:Traj_ODFal2}
\end{figure*}

\paragraph{Evaluation using \textsc{DTFal}} The reported falsification time includes the time to train the decision tree since decision tree learning is specification dependent. The algorithm is invoked with a command to generate one counterexample and allowing maximum 1 retraining. The size of the dataset  and the number of random simulations to search from an explanation are chosen on a trail and error basis, with the goal of keeping it as small as possible and at the same time succeeding in falsification. We observe that \textsc{DTFal} shows high falsification rate in most of the instances and more importantly, it displays high success rate, falsifying $33$ out of $37$ instances. The falsification time in \textsc{DTFal} is better than \textsc{NNFal} in most instances. Overall, we can conclude that \textsc{DTFal} is more effective than \textsc{NNFal} in our experiments. Figure \ref{fig:Traj_NNFal} depicts counterexamples generated by \textsc{NNFal} on Oscillator, Navigation, and ACC. Figure \ref{fig:Traj_ODFal1} and \ref{fig:Traj_ODFal2} show the counterexamples obtained by \textsc{DTFal} on some of the benchmark instances.

\subsection{\textbf{RQ2} Comparison with CPS falsification tools}

\noindent We present a comparison of \textsc{FlexiFal} with the falsification tools that participated in ARCH-COMP 24~\cite{khandait2024arch}. The performance of these tools is shown in Table \ref{tab:fal_table}. The data shown in the table is from the ARCH-COMP 2024~\cite{khandait2024arch} report, except the appended results of \textsc{FlexiFal}. Participating tools use different methodologies, making it challenging to decide on the comparison metrics and judging the best-performing tool. Over the past editions of the competition, falsification rate (FR) along with the mean number of simulations ($\overline{S}$) required to falsify an instance are the agreed upon metrics that each participating tool report. The primary objective of the competition is to report the state-of-the-art falsification tools, highlighting their strengths and limitations for different benchmarks. We observe that \textsc{FReak} requires a very small number of simulations to falsify the instances among the other tools. However, it cannot falsify F16 and AFC33, where \textsc{DTFal} succeeds. The falsification rate (FR) is nearly the same for all the instances among the tools. The mean number of simulations needed to falsify an instance varies considerably across the tools. We see that \textsc{DTFal} algorithm of \textsc{FlexiFal} finds counterexamples in most of the instances with high FR rate. However, it generally requires a larger number of simulations than the others. On the brighter side, \textsc{DTFal} identifies counterexamples in instances such as F16 and AFC33, which most of the other tools fail to falsify. We therefore believe that \textsc{FlexiFal} is competent among the state-of-the-art.

\begin{table*}[htbp]
    \caption{A Comparison with the CPS falsification tools. FR is the falsification rate. $\overline{S}$ is the mean number of simulations needed to falsify the respective instances. The blank in the table indicates that the respective tools could not falsify the instances.}
    \centering
    % \footnotesize
    \begin{adjustbox}{width=1\textwidth}
    \begin{tabular}{|c|c|c|c|c|c|c|c|c|c|c|c|c|c|c|c|c|c|c|c|c|c|c|c|c|}
         \hline
         Tool & \multicolumn{2}{c|}{UR} & \multicolumn{2}{c|}{ARIsTEO} & \multicolumn{2}{c|}{ATheNA} & \multicolumn{2}{c|}{EXAM-Net} & \multicolumn{2}{c|}{FalCAun} & \multicolumn{2}{c|}{ForeSee} & \multicolumn{2}{c|}{FReak} & \multicolumn{2}{c|}{Moonlight} & \multicolumn{2}{c|}{FlexiFal} & \multicolumn{2}{c|}{OD} & \multicolumn{2}{c|}{$\Psi-TaLiRo$} & \multicolumn{2}{c|}{$\Psi-TaLiRo$} \\
         Approach & \multicolumn{2}{c|}{} & \multicolumn{2}{c|}{ARX-2} & \multicolumn{2}{c|}{} & \multicolumn{2}{c|}{} & \multicolumn{2}{c|}{} & \multicolumn{2}{c|}{} & \multicolumn{2}{c|}{} & \multicolumn{2}{c|}{} & \multicolumn{2}{c|}{(DTFal)} & \multicolumn{2}{c|}{} & \multicolumn{2}{c|}{ConBo-LS} & \multicolumn{2}{c|}{PART-X} \\
         \hline
         Instance & FR & $\overline{s}$ & FR & $\overline{s}$ & FR & $\overline{s}$ & FR & $\overline{s}$ & FR & $\overline{s}$ & FR & $\overline{s}$ & FR & $\overline{s}$ & FR & $\overline{s}$ & FR & $\overline{s}$ & FR & $\overline{s}$ & FR & $\overline{s}$ & FR & $\overline{s}$ \\
         \hline
         F16 & & & & & & & & & & & & & & & 10 & 21 & 10 & 371.72 & & & & & & \\
         \hline
         CC1 & 10 & 16.4 & 10 & 11.3 & 10 & 60.5 & 10 & 36.1 & 10 & 396 & 10 & 21.8 & 10 & 3 & & & 10 & 169.90 & 10 & 61.2 & 10 & 18 & 10 & 17.6\\
         \hline
         CC2 & 10 & 12.4 & 10 & 10.5 & 10 & 94 & 9 & 398.6 & 10 & 76 & 8 & 224.3 & 10 & 3 & & & 10 & 241.09 & 4 & 109.3 & 10 & 14.8 & 10 & 17.8 \\
         \hline
         CC3 & 10 & 19.6 & 10 & 18.8 & 10 & 119.2 & 10 & 14.9 & 10 & 122 & 10 & 36.1 & 10 & 2.6 & & & 10 & 152.1 & 10 & 35.2 & 10 & 11.4 & 10 & 13.5 \\
         \hline
         CC4 & & & & & 2 & 514 & & & 4 & 1256 & 7 & 680 & 10 & 1349.9 & & & & & & & 1 & 1387 & & \\
         \hline
         CC5 & 10 & 37.4 & 10 & 29.1 & 9 & 112 & 10 & 78.2 & & & 10 & 88.7 & 10 & 47.5 & & & 10 & 523.1 & 10 & 55.7 & 10 & 32.3 & 10 & 29.9\\
         \hline
         CCx & 6 & 396.7 & 9 & 610.4 & 5 & 86.4 & 10 & 448.1 & & & 10 & 228 & 7 & 1723.6 & & & 10 & 468.66 & 2 & 261.5 & 10 & 244.2 & 10 & 607 \\
         \hline
         AT1 & & & & & & & 10 & 150.4 & 10 & 896 & 10 & 387.4 & 10 & 4.5 & & & 10 & & & & 5 & 1131.7 & 10 & 30.5 \\
         \hline
         AT2 & 10 & 18.8 & 10 & 15.1 & 10 & 62.3 & 10 & 22.7 & 10 & 256 & 9 & 196.6 & 10 & 2.2 & & & 10 & 51 & 10 & 18.5 & 10 & 13.7 & 10 & 6.5 \\
         \hline
         AT51 & 10 & 20.5 & 1 & 1371 & 10 & 106 & 10 & 8.5 & & & 10 & 15 & 10 & 17.7 & & & 10 & 57.78 & 10 & 11.6 & 10 & 10.1 & 10 & 13.3 \\
         \hline
         AT52 & 10 & 74.1 & 10 & 4.4 & 10 & 19 & 10 & 10.3 & & & 10 & 65.7 & 10 & 5.2 & & & 10 & 52.8 & 10 & 8.6 & 10 & 67.8 & 10 & 66.5\\
         \hline
         AT53 & 10 & 1.5 & 10 & 4.4 & 10 & 2.2 & 10 & 1.5 & & & 10 & 4.3 & 10 & 3.9 & & & 10 & 53.1 & 10 & 2.2 & 10 & 4.1 & 10 & 2.2 \\
         \hline
         AT54 & 10 & 47.9 & 6 & 571.5 & 10 & 138.9 & 10 & 89.6 & & & 10 & 60.3 & 10 & 103 & & & 10 & & 10 & 18.4 & 10 & 37.4 & 10 & 85 \\
         \hline
         AT6a & 10 & 156.6 & 8 & 271.4 & 10 & 245.5 & 10 & 61.1 & 10 & 1002 & 10 & 115.2 & 10 & 24.6 & & & 10 & 197.9 & 10 & 64.3 & 10 & 304.5 & 10 & 153.7 \\
         \hline
         AT6b & 10 & 472.2 & 6 & 536 & 9 & 279.3 & 10 & 119.4 & & & 10 & 253.9 & 10 & 17.4 & & & 10 & 314.2 & 10 & 122 & 6 & 782.2 & 10 & 307.9 \\
         \hline
         AT6c & 10 & 326.8 & 8 & 643.8 & 10 & 194.5 & 10 & 176.5 & 10 & 898 & 10 & 133.1 & 10 & 15.5 & & & 10 & 292.2 & 10 & 118.5& 9 & 472.1 & 10 & 334.4 \\
         \hline
         AT6abc & 10 & 149 & 8 & 505.2 & 10 & 234.4 & 10 & 57.4 & 10 & 1232 & 10 & 123.6 & 10 & 13.1 & & & 10 & 220.8 & 10 & 61.3 & 10 & 139.2 & 10 & 106.9 \\
         \hline
         AFC27 & & & 9 & 56.2 & 8 & 137.8 & 10 & 235.8 & & & & & 10 & 12 & & & 10 & 25.7 & 10 & 391.2 & 10 & 101.8 & 10 & 34.3 \\
         \hline
         AFC29 & 10 & 25.1 & 10 & 3.9 & 10 & 18.6 & 10 & 8.5 & & & & & 10 & 3.1 & & & 10 & 21 & 10 & 13 & 10 & 9.1 & 10 & 12.1 \\
         \hline
         AFC33 & & & & & & & & & & & & & & & & & 10 & 21 & & & & & & \\
         \hline
         
    \end{tabular}
    \end{adjustbox}
    \label{tab:fal_table}
\end{table*}

\begin{figure}[h]
    \centering
    \includegraphics[width=\linewidth]{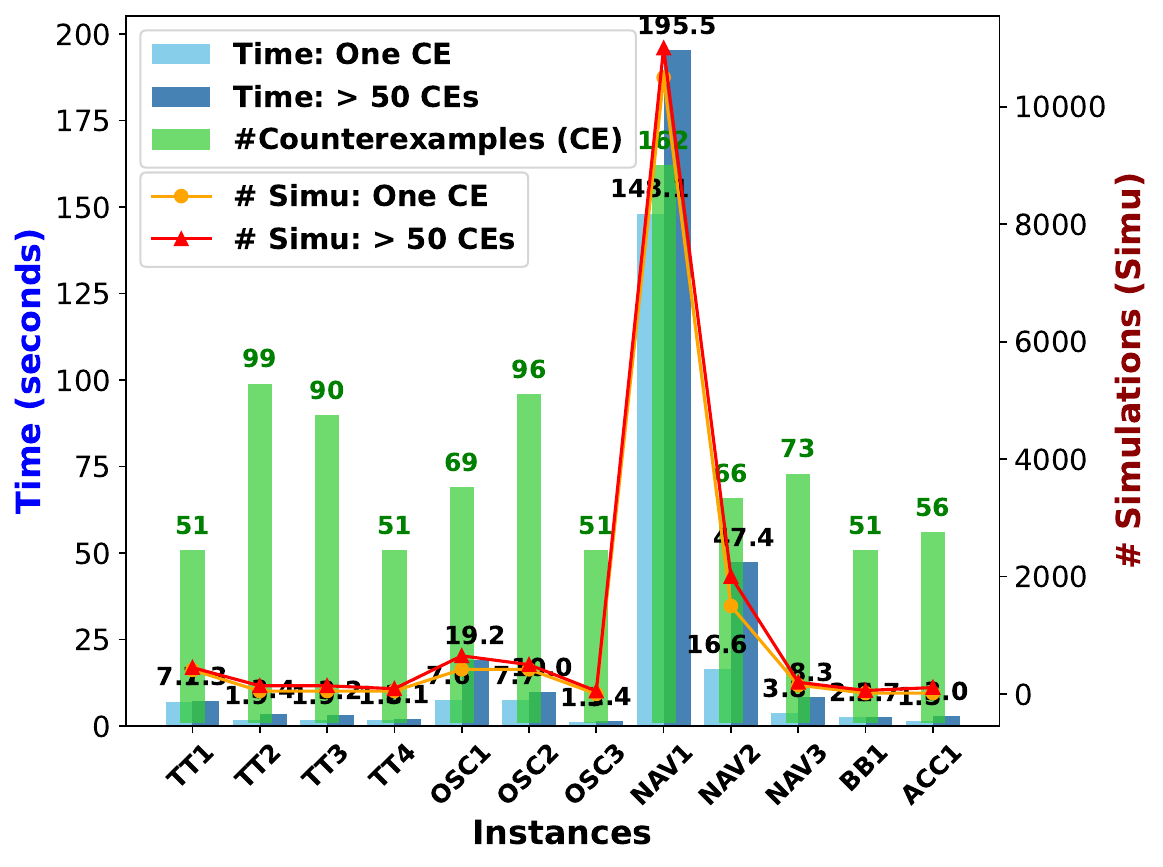}
    \caption{\#Simulations to find one vs multiple counterexamples nearly overlap. The time difference between finding one counterexample vs many is small in most of the instances.}
    \label{fig:onevsmanyha}
\end{figure}

\begin{figure}[h]
    \centering
    \includegraphics[width=\linewidth]{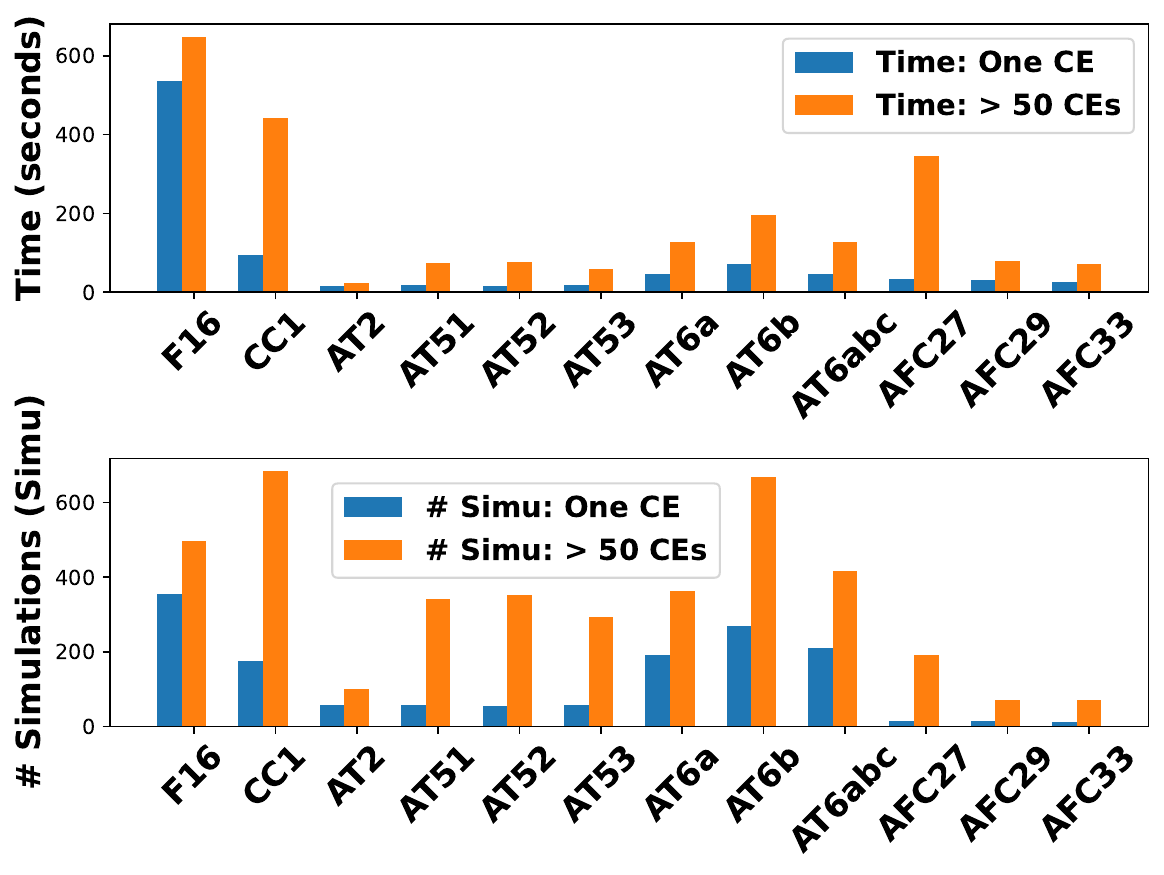}
    \caption{Top: Time for finding 1 vs more than 50 counterexamples. Bottom: \#simulations to find 1 vs more than 50 counterexamples across the Simulink instances.}
    \label{fig:onevsmanysimu}
\end{figure}

\subsection{\textbf{RQ3:} Generating multiple counterexamples efficiently}
This section evaluates the ability of \textsc{FlexiFal} to generate multiple counterexamples efficiently. The \textsc{NNFal} algorithm relies primarily on the DNN verifier or adversarial attack algorithms, which requires considerable computational effort just to obtain a single counterexample. Running the algorithm multiple times to find new counterexamples is expensive.  In contrast, the \textsc{DTFal} algorithm builds explanations for property violation, which can be potentially used to find multiple counterexamples with a little extra computational effort. The computational effort is measured by the number of random simulations and the time required to find them. Figures \ref{fig:onevsmanyha} and \ref{fig:onevsmanysimu} show the results of finding many ($>$ 50) counterexamples, and we see that \textsc{DTFal} finds them with a little extra computational effort compared to finding just one.

\noindent Figure \ref{fig:onevsmanysimu} shows the comparison for instances on Simulink models. 
For example, \textsc{DTFal} finds a single counterexample in F16 using $400$ simulations in $513$ seconds. In comparison, \textsc{DTFal}  generates $50$ counterexamples using $500$ simulations in $650$ seconds, that is 49 more in less than 150 seconds with just 100 extra simulations.

% \vspace{-5mm}
\section{Related Works}
 \noindent \textbf{Surrogate model-based CPS falsification:} Recently,~\cite{DBLP:conf/icse/MenghiNBP20, waga2020falsification} propose learning surrogate models from the CPS executions aiming to use them for falsification. A black-box technique with an approximation-refinement strategy is reported in \cite{DBLP:conf/icse/MenghiNBP20} in which compute-intensive CPS that can take hours to simulate is converted to a surrogate model that is faster to execute. Surrogate models such as the Hammerstein-Wiener or the non-linear ARX model are generated using Matlab's SI (System Identification) toolbox from the system's input-output response. The surrogate is then tested against the specifications represented in signal temporal logic (STL). A falsifying trajectory is validated on the original CPS, and the surrogate model is iteratively refined to eliminate spurious trajectories. A robustness-guided black box checking for falsification of STL properties is reported in \cite{waga2020falsification}, in which a Mealy machine is constructed using automata learning. The counterexample in the Mealy machine is validated, and the model is refined to eliminate spurious counterexamples. \textsc{FReak}~\cite{bak2024hscc}  is a data-driven falsification tool that employs a surrogate model to replicate system dynamics. It uses Koopman operator linearization to construct a linear model of non-linear dynamics systems as a surrogate. The tool uses reachable analysis of the surrogate model and uses the reachability knowledge in the encoding of STL specifications into a Mixed-Integer Linear Program (MILP) to identify the least robust trajectory within the reachable state-space. If the trajectory is spurious, additional simulation data is generated to retrain the Koopman model and repeat the process until the instance is falsified. None of these works use machine learning models as CPS surrogates, which makes them different from our work.

\noindent \textbf{Falsification tools for Simulink models:} On the other hand, several state-of-the-art falsification tools for CPS represented as Simulink models are reported in ~\cite{ernst2019fast, donze2010breach, DBLP:conf/fmics/ThibeaultACPF21, DBLP:conf/icse/MenghiNBP20, annpureddy2011s, waga2020falsification}. These algorithms are simulation-based falsification procedures for specifications written in Metric Temporal Logic (MTL) or Signal Temporal Logic (STL). A falsification algorithm proposed in~\cite{ernst2019fast} is based on adaptive Las-Vegas Tree Search (aLVTS), which constructs falsifying inputs incrementally in time. The fundamental concept is to start with simple inputs and scale up gradually by taking samples from input domains with improved temporal and spatial resolution. In~\cite{annpureddy2011s}, a toolbox is reported for falsifying MTL properties based on stochastic optimization techniques such as Monte-Carlo methods and Ant-Colony Optimization. The stochastic sampler suggests an input signal to the simulator which returns a trajectory. The robustness analyzer then examines the trajectory and returns a robustness value. The negative robustness value indicates that the toolbox falsifies the temporal property. A positive robustness value is used by the sampler to decide the next input. $\Psi$-TaLiRo~\cite{DBLP:conf/fmics/ThibeaultACPF21} is another robustness-guided falsification tool for CPS, which is a Python implementation of the toolbox proposed in \cite{annpureddy2011s}. The updated version supports built-in optimizers like DA and Uniform Random, with two new optimization algorithms such as Conjunctive Bayesian Optimization - Large Scale (ConBO-LS) and Part-X. 

Breach~\cite{donze2010breach} is a MATLAB toolbox mainly designed for STL specification monitoring and test case generation for hybrid dynamical systems. In addition, it supports optimization-based falsification and requirements mining for CPS. In contrary to these works, our falsification framework is data driven, making it application to any executable CPS in general and not limited to Simulink models per se.

 \noindent \textbf{Falsification tools for Hybrid Automaton Models:} Hybrid automaton models of CPS have been studied with the focus on verification and model-checking. Some of the model-checkers for HA, such as \textsc{dReach} \cite{kong2015dreach}, \textsc{Hylaa} \cite{DBLP:conf/hybrid/BakD17}, \textsc{XSpeed} \cite{DBLP:conf/hvc/RayGDBBG15},  \textsc{SAT-Reach}~\cite{10.1145/3567425}), have counterexample generation feature for reachability specifications expressed in their tool-specific language. As the computational effort in these tools is directed towards symbolic state-space exploration for proving safety rather than falsification, the computational effort that these tools incur for falsification is not comparable with falsification methods and tools. However, \textsc{FlexiFal} is capable of efficiently falsifying HA models with STL specifications. 

\section{Conclusion}
\noindent A data-driven framework has been proposed to falsify CPS safety specifications, which is primarily based on the building of a surrogate model, approximating a CPS given as an executable. The framework has the option to construct either a feedforward neural network or a decision tree as a surrogate model. The network-based falsification algorithm leverages \emph{adversarial attack} algorithms and efficient verification tools for the falsification of reachability specifications. The use of a decision tree as a surrogate model has shown great promise on the falsification of CPS. The experimental evaluation indicates that our framework \textsc{FlexiFal} finds multiple hard-to-find counterexamples in CPS. The datasets and network surrogates of CPS have been made available in the public domain, which can be useful for novel machine learning-based analysis algorithms for CPS. 

\bibliographystyle{plain}

\begin{thebibliography}{10}

\bibitem{Alur92hybridautomata}
Rajeev Alur, Costas Courcoubetis, Thomas~A. Henzinger, and Pei-Hsin Ho.
\newblock Hybrid automata: An algorithmic approach to the specification and verification of hybrid systems.
\newblock pages 209--229. Springer-Verlag, 1992.

\bibitem{annpureddy2011s}
Yashwanth Annpureddy, Che Liu, Georgios Fainekos, and Sriram Sankaranarayanan.
\newblock S-taliro: A tool for temporal logic falsification for hybrid systems.
\newblock In {\em TACAS 2011, Saarbr{\"u}cken, Germany, March 26--April 3, 2011.}, pages 254--257. Springer, 2011.

\bibitem{bak2024hscc}
Stanley Bak, Sergiy Bogomolov, Abdelrahman Hekal, Niklas Kochdumper, Ethan Lew, Andrew Mata, and Amir Rahmati.
\newblock Falsification using reachability of surrogate koopman models.
\newblock In {\em HSCC 24}, pages 1--13, 2024.

\bibitem{DBLP:conf/hybrid/BakD17}
Stanley Bak and Parasara~Sridhar Duggirala.
\newblock Hylaa: {A} tool for computing simulation-equivalent reachability for linear systems.
\newblock In {\em HSCC, Pittsburgh, PA, USA, April 18-20, 2017}, pages 173--178. {ACM}, 2017.

\bibitem{bak2020improved}
Stanley Bak, Hoang-Dung Tran, Kerianne Hobbs, and Taylor~T Johnson.
\newblock Improved geometric path enumeration for verifying relu neural networks.
\newblock In {\em CAV 2020, Los Angeles, CA, USA, July 21--24, 2020}, pages 66--96. Springer, 2020.

\bibitem{breiman2017classification}
Leo Breiman.
\newblock {\em Classification and regression trees}.
\newblock Routledge, 2017.

\bibitem{DBLP:conf/arch/BuAAMRWZ20}
Lei Bu, Alessandro Abate, Dieky Adzkiya, Muhammad~Syifa'ul Mufid, Rajarshi Ray, Yuming Wu, and Enea Zaffanella.
\newblock {ARCH-COMP20} category report: Hybrid systems with piecewise constant dynamics and bounded model checking.
\newblock In {\em {ARCH 2020.}}, volume~74 of {\em EPiC Series in Computing}, pages 1--15. EasyChair, 2020.

\bibitem{DBLP:conf/sp/Carlini017}
Nicholas Carlini and David~A. Wagner.
\newblock Towards evaluating the robustness of neural networks.
\newblock In {\em 2017 {IEEE} Symposium on Security and Privacy, {SP} 2017, San Jose, CA, USA, May 22-26, 2017}, pages 39--57. {IEEE} Computer Society, 2017.

\bibitem{donze2010breach}
Alexandre Donz{\'e}.
\newblock Breach, a toolbox for verification and parameter synthesis of hybrid systems.
\newblock In {\em CAV 2010, Edinburgh, UK, July 15-19, 2010. Proceedings 22}, pages 167--170. Springer, 2010.

\bibitem{Donze2010Robust}
Alexandre Donz{\'e} and Oded Maler.
\newblock Robust satisfaction of temporal logic over real-valued signals.
\newblock In Krishnendu Chatterjee and Thomas~A. Henzinger, editors, {\em Formal Modeling and Analysis of Timed Systems}, pages 92--106, Berlin, Heidelberg, 2010. Springer Berlin Heidelberg.

\bibitem{ernst2019fast}
Gidon Ernst, Sean Sedwards, Zhenya Zhang, and Ichiro Hasuo.
\newblock Fast falsification of hybrid systems using probabilistically adaptive input.
\newblock In {\em QEST 2019, Glasgow, UK, September 10--12, 2019, Proceedings 16}, pages 165--181. Springer, 2019.

\bibitem{FanQM017}
Chuchu Fan, Bolun Qi, Sayan Mitra, and Mahesh Viswanathan.
\newblock Dryvr: Data-driven verification and compositional reasoning for automotive systems.
\newblock In {\em {CAV} 2017, Heidelberg, Germany, July 24-28, 2017, Proceedings, Part {I}}, volume 10426, pages 441--461. Springer, 2017.

\bibitem{FehnkerI04}
Ansgar Fehnker and Franjo Ivancic.
\newblock Benchmarks for hybrid systems verification.
\newblock In {\em HSCC}, pages 326--341, 2004.

\bibitem{FLGDCRLRGDM11}
Goran Frehse, Colas Le~Guernic, Alexandre Donz\'e, Scott Cotton, Rajarshi Ray, Olivier Lebeltel, Rodolfo Ripado, Antoine Girard, Thao Dang, and Oded Maler.
\newblock {SpaceEx: Scalable Verification of Hybrid Systems}.
\newblock In {\em CAV}, LNCS. Springer, 2011.

\bibitem{goodfellow2014explaining}
Ian~J Goodfellow, Jonathon Shlens, and Christian Szegedy.
\newblock Explaining and harnessing adversarial examples.
\newblock {\em arXiv preprint arXiv:1412.6572}, 2014.

\bibitem{DBLP:conf/adhs/HeidlaufCBB18}
Peter Heidlauf, Alexander Collins, Michael Bolender, and Stanley Bak.
\newblock Verification challenges in {F-16} ground collision avoidance and other automated maneuvers.
\newblock In {\em ARCH@ADHS 2018, Oxford, UK, July 13, 2018}, volume~54, pages 208--217. EasyChair, 2018.

\bibitem{DBLP:conf/hicss/Hiskens01}
Ian~A. Hiskens.
\newblock Stability of limit cycles in hybrid systems.
\newblock In {\em 34th Annual Hawaii International Conference on System Sciences (HICSS-34), January 3-6, 2001, Maui, Hawaii, {USA}}. {IEEE} Computer Society, 2001.

\bibitem{DBLP:conf/cpsweek/HoxhaAF14}
Bardh Hoxha, Houssam Abbas, and Georgios Fainekos.
\newblock Benchmarks for temporal logic requirements for automotive systems.
\newblock In {\em ARCH@CPSWeek 2014, Berlin, Germany, April 14, 2014 / ARCH@CPSWeek 2015, Seattle, WA, USA, April 13, 2015}, volume~34, pages 25--30. EasyChair, 2014.

\bibitem{10.5555/646880.710449}
Jianghai Hu, John Lygeros, and Shankar Sastry.
\newblock Towars a theory of stochastic hybrid systems.
\newblock In {\em HSCC 2000}, page 160–173, Berlin, Heidelberg, 2000. Springer-Verlag.

\bibitem{jin2014powertrain}
Xiaoqing Jin, Jyotirmoy~V Deshmukh, James Kapinski, Koichi Ueda, and Ken Butts.
\newblock Powertrain control verification benchmark.
\newblock In {\em Proceedings of the 17th international conference on Hybrid systems: computation and control}, pages 253--262, 2014.

\bibitem{katz2017reluplex}
Guy Katz, Clark Barrett, David~L Dill, Kyle Julian, and Mykel~J Kochenderfer.
\newblock Reluplex: An efficient smt solver for verifying deep neural networks.
\newblock In {\em CAV 2017, Heidelberg, Germany, July 24-28, 2017, Proceedings, Part I 30}, pages 97--117. Springer, 2017.

\bibitem{khandait2024arch}
Tanmay Khandait, Federico Formica, Paolo Arcaini, Surdeep Chotaliya, Georgios Fainekos, Abdelrahman Hekal, Atanu Kundu, Ethan Lew, Michele Loreti, Claudio Menghi, et~al.
\newblock Arch-comp 2024 category report: Falsification.
\newblock In {\em Proceedings of the 11th Int. Workshop on Applied}, volume 103, pages 122--144, 2024.

\bibitem{kong2015dreach}
Soonho Kong, Sicun Gao, Wei Chen, and Edmund Clarke.
\newblock dreach: $\delta$-reachability analysis for hybrid systems.
\newblock In {\em TACAS 2015, London, UK, April 11-18, 2015, Proceedings}, volume 9035, page 200. Springer, 2015.

\bibitem{10.1145/3567425}
Atanu Kundu, Sarthak Das, and Rajarshi Ray.
\newblock Sat-reach: A bounded model checker for affine hybrid systems.
\newblock {\em ACM Trans. Embed. Comput. Syst.}, 22(2), jan 2023.

\bibitem{DBLP:conf/iclr/KurakinGB17a}
Alexey Kurakin, Ian~J. Goodfellow, and Samy Bengio.
\newblock Adversarial examples in the physical world.
\newblock In {\em 5th International Conference on Learning Representations, {ICLR} 2017, Toulon, France, April 24-26, 2017, Workshop Track Proceedings}. OpenReview.net, 2017.

\bibitem{madry2017towards}
Aleksander Madry, Aleksandar Makelov, Ludwig Schmidt, Dimitris Tsipras, and Adrian Vladu.
\newblock Towards deep learning models resistant to adversarial attacks.
\newblock {\em arXiv preprint arXiv:1706.06083}, 2017.

\bibitem{maler2004monitoring}
Oded Maler and Dejan Nickovic.
\newblock Monitoring temporal properties of continuous signals.
\newblock In {\em International Symposium on Formal Techniques in Real-Time and Fault-Tolerant Systems}, pages 152--166. Springer, 2004.

\bibitem{DBLP:conf/icse/MenghiNBP20}
Claudio Menghi, Shiva Nejati, Lionel~C. Briand, and Yago~Isasi Parache.
\newblock Approximation-refinement testing of compute-intensive cyber-physical models: an approach based on system identification.
\newblock In Gregg Rothermel and Doo{-}Hwan Bae, editors, {\em {ICSE} '20, Seoul, South Korea, 27 June - 19 July, 2020}, pages 372--384. {ACM}, 2020.

\bibitem{DBLP:conf/cvpr/Moosavi-Dezfooli16}
Seyed{-}Mohsen Moosavi{-}Dezfooli, Alhussein Fawzi, and Pascal Frossard.
\newblock Deepfool: {A} simple and accurate method to fool deep neural networks.
\newblock In {\em {CVPR} 2016, Las Vegas, NV, USA, June 27-30}, pages 2574--2582. {IEEE} Computer Society, 2016.

\bibitem{DBLP:conf/hvc/RayGDBBG15}
Rajarshi Ray, Amit Gurung, Binayak Das, Ezio Bartocci, Sergiy Bogomolov, and Radu Grosu.
\newblock {XSpeed: Accelerating Reachability Analysis on Multi-core Processors}.
\newblock In {\em {HVC} 2015, Haifa, Israel, November 17-19, Proceedings}, pages 3--18, 2015.

\bibitem{rony2019decoupling}
J{\'e}r{\^o}me Rony, Luiz~G Hafemann, Luiz~S Oliveira, Ismail~Ben Ayed, Robert Sabourin, and Eric Granger.
\newblock Decoupling direction and norm for efficient gradient-based l2 adversarial attacks and defenses.
\newblock In {\em Proceedings of the IEEE/CVF Conference on Computer Vision and Pattern Recognition}, pages 4322--4330, 2019.

\bibitem{serban2005cvodes}
Radu Serban and Alan~C Hindmarsh.
\newblock Cvodes: the sensitivity-enabled ode solver in sundials.
\newblock In {\em International Design Engineering Technical Conferences and Computers and Information in Engineering Conference}, volume 47438, pages 257--269, 2005.

\bibitem{shriver2021reducing}
David Shriver, Sebastian Elbaum, and Matthew~B Dwyer.
\newblock Reducing dnn properties to enable falsification with adversarial attacks.
\newblock In {\em 2021 IEEE/ACM 43rd International Conference on Software Engineering (ICSE)}, pages 275--287. IEEE, 2021.

\bibitem{DBLP:conf/cav/ShriverED21}
David Shriver, Sebastian~G. Elbaum, and Matthew~B. Dwyer.
\newblock {DNNV:} {A} framework for deep neural network verification.
\newblock In {\em {CAV} 2021, July 20-23, Proceedings, Part {I}}, volume 12759, pages 137--150. Springer, 2021.

\bibitem{DBLP:journals/corr/SzegedyZSBEGF13}
Christian Szegedy, Wojciech Zaremba, Ilya Sutskever, Joan Bruna, Dumitru Erhan, Ian~J. Goodfellow, and Rob Fergus.
\newblock Intriguing properties of neural networks.
\newblock In {\em {ICLR} 2014, Banff, AB, Canada, April 14-16, Conference Track Proceedings}, 2014.

\bibitem{DBLP:conf/fmics/ThibeaultACPF21}
Quinn Thibeault, Jacob Anderson, Aniruddh Chandratre, Giulia Pedrielli, and Georgios Fainekos.
\newblock Psy-taliro: {A} python toolbox for search-based test generation for cyber-physical systems.
\newblock In {\em {FMICS} 2021, Paris, France, August 24-26, Proceedings}, volume 12863, pages 223--231. Springer, 2021.

\bibitem{waga2020falsification}
Masaki Waga.
\newblock Falsification of cyber-physical systems with robustness-guided black-box checking.
\newblock In {\em Proceedings of HSCC 23}, pages 1--13, 2020.

\bibitem{wang2018efficient}
Shiqi Wang, Kexin Pei, Justin Whitehouse, Junfeng Yang, and Suman Jana.
\newblock Efficient formal safety analysis of neural networks.
\newblock {\em Advances in neural information processing systems}, 31, 2018.

\bibitem{yan2025automated}
Yipei Yan, Deyun Lyu, Zhenya Zhang, Paolo Arcaini, and Jianjun Zhao.
\newblock Automated generation of benchmarks for falsification of stl specifications.
\newblock {\em IEEE Transactions on Computer-Aided Design of Integrated Circuits and Systems}, 2025.

\bibitem{yuan2019adversarial}
Xiaoyong Yuan, Pan He, Qile Zhu, and Xiaolin Li.
\newblock Adversarial examples: Attacks and defenses for deep learning.
\newblock {\em IEEE transactions on neural networks and learning systems}, 30(9):2805--2824, 2019.

\end{thebibliography}

\end{document}